\documentclass[usenatbib,usegraphicx]{mn2e}
\usepackage{array}
\usepackage{url}
\usepackage{graphicx}
\usepackage{fleqn}
\usepackage{appendix}
\usepackage{amsmath}
\usepackage{lineno}

\bibpunct{(}{)}{;}{a}{}{,}

\newcommand{\ped}{period echelle diagram}
\newcommand{\kep}{{\it Kepler}}

\newcommand{\kicnineone}{KIC\,9145955}
\newcommand{\kicfiveeight}{KIC\,5866737}

\newcommand{\kiconezero}{KIC\,10200377}

\newcommand{\blank}{\ensuremath{\cdots}}
\newcommand{\msun}{\ensuremath{M_\odot}}

\newcommand{\logg}{\ensuremath{\log g}}
\newcommand{\muHz}{\ensuremath{\mu\mathrm{Hz}}}
\newcommand{\numax}{\ensuremath{\nu_\mathrm{max}}}
\newcommand{\logtg}{$\log T_\mathrm{eff}$--$\log g$}
\newcommand{\dpg}{\ensuremath{\Delta \Pi_{1,\mathrm{g}}}}
\newcommand{\dplg}{\ensuremath{\Delta \Pi_{l,\mathrm{g}}}}
\newcommand{\dpl}{\ensuremath{\Delta \Pi_l}}
\newcommand{\pir}{\ensuremath{\Pi_\mathrm{r}}}
\newcommand{\pirg}{\ensuremath{\pir^{(\mathrm{g})}}}
\newcommand{\pirgi}{\ensuremath{\pir^{(\mathrm{g},i)}}}
\newcommand{\pirpi}{\ensuremath{\pir^{(\mathrm{p},i)}}}
\newcommand{\pirir}{\ensuremath{\pir^{(i,\mathrm{r})}}}
\newcommand{\piril}{\ensuremath{\pir^{(i,\mathrm{l})}}}
\newcommand{\nuir}{\ensuremath{\nu_{l=1}^{(i,\mathrm{r})}}}
\newcommand{\nuil}{\ensuremath{\nu_{l=1}^{(i,\mathrm{l})}}}

\newcommand{\chisq}{\ensuremath{\chi^2}}
\newcommand{\chisqvert}{\ensuremath{\chisq_\mathrm{vert}}}
\newcommand{\chisqsymm}{\ensuremath{\chisq_\mathrm{symm}}}
\newcommand{\chisqtot}{\ensuremath{\chisq_\mathrm{tot}}}

\def\apj{ApJ}%
\def\apjl{ApJ}%
\def\apjs{ApJS}%
\def\apss{Ap\&SS}%
\def\aap{A\&A}%
\def\mnras{MNRAS}%
\def\pasp{PASP}%
\def\pasj{PASJ}%
\def\nat{Nature}%

\def\myfigure#1#2#3#4{
        \begin{figure#4}
	\centering
        \resizebox{\hsize}{!}{\includegraphics{#1}}
        \caption{#2 \label{#3}}
        \end{figure#4}
}

\def\myfiguretwobyone#1#2#3#4#5{
        \begin{figure#5}
	\centering
        \resizebox{\hsize}{!}{\includegraphics{#1}}
        \resizebox{\hsize}{!}{\includegraphics{#2}}
        \caption{#3 \label{#4}}
        \end{figure#5}
}

\begin{document}

\title{Automated determination of g-mode period spacing of red-giant stars}
\author[A.~Datta et al.]{%
Abhisek Datta$^1$,
Anwesh Mazumdar$^2$,
Umang Gupta$^3$,
and
Saskia Hekker$^{4,5,6}$
\\
\\
$^1$Department of Physics, 
Indian Institute of Technology Kharagpur, 
Kharagpur 721302, India.\\
$^2$Homi Bhabha Centre for Science Education, TIFR,
V.~N.~Purav Marg, Mankhurd, Mumbai 400088, India.\\
$^3$Department of Electrical Engineering,
Indian Institute of Technology Delhi,
Hauz Khas, New Delhi 110016, India.\\
$^4$Astronomical Institute `Anton Pannenkoek', University of Amsterdam, 
Science Park 904, 1098 HX Amsterdam, the Netherlands.\\
$^5$Max Planck Institut f\"ur Sonnensystemforschung, 
Justus-von-Liebig-Weg 3, 37077 G\"ottingen, Germany.\\
$^6$Stellar Astrophysics Centre, Department of Physics and Astronomy, 
Aarhus University, Ny Munkegade 120, DK-8000 Aarhus C,\\
Denmark.
}%

\maketitle

\begin{abstract}

The \kep\ satellite has provided photometric timeseries data of
unprecedented length, duty cycle and precision. To fully analyse these
data for the tens of thousands of stars observed by \kep, automated
methods are a prerequisite. Here we present an automated procedure to
determine the period spacing of gravity modes in red-giant stars
ascending the red-giant branch. The gravity modes reside in a cavity in
the deep interior of the stars and provide information on the conditions
in the stellar core. However, for red giants the gravity modes are not
directly observable on the surface, hence this method is based on the
pressure-gravity mixed modes that present observable features in the
Fourier power spectrum. The method presented here is based on the
vertical alignment and symmetry of these mixed modes in a period echelle
diagram.  We find that we can obtain reliable results for both model
frequencies and observed frequencies. Additionally, we carried out Monte
Carlo tests to obtain realistic uncertainties on the period spacings
with different set of oscillation modes (for the models) and
uncertainties on the frequencies. Furthermore, this method has been
used to improve mode detection and identification of the observed
frequencies in an iterative manner.

\end{abstract}

\section{Introduction}
\label{sec:intro}

Red giants are evolved stars which have a hydrogen-depleted core
surrounded by a hydrogen-burning shell. Further evolved red giants
also burn helium in their core. Although the structure of red giants
is broadly understood from the theory of stellar structure and
evolution, several important questions remain, such as the mechanism
of convective heat flow, nuclear processes in extremely dense
material, and transport of angular momentum through differential
rotation. Several recent studies have indeed focussed on these
questions using the techniques of asteroseismology which is the study
of stellar oscillation frequencies.

Among the recent results obtained are the detection of signatures from
the core through so-called mixed oscillation modes
\citep{Beck11,Bedding11,Mosser12,Mosser14} and theoretical explanations of
these modes \citep{JCD08,Dupret09,Montalban10,JCD11}. These mixed modes are
subsequently used to probe differential rotation in the stars
\citep{Beck12,Deheuvels12,Mosser12rot,Deheuvels14}. Efforts to fully
incorporate rotation and angular momentum transport in stellar
evolution models have been made, but these results do not yet provide
results consistent with observations
\citep{Ceillier13,Goupil13,Marques13,Ouazzani13,Goupil14,Cantiello14}.

In general, there are two main sets of solutions for the equation of
motion of a pulsating star and these lead to two types of pulsation
modes: p-modes and g-modes (for a detailed discussion on these, see
\citet{Chaplin13} and \citet{Hekker13}).  Due to the large density
gradient outside the helium core, a red giant is effectively divided
into two cavities. In the envelope, the non-radial oscillations behave
as p-modes, while in the core they behave like g-modes. The models
predict a very dense spectrum of these so-called mixed modes for each
value of $l$ (except $l=0$). Details about mixed modes and resonant
coupling between the two cavities in red giants have been described by
\citet{Bedding11} and \citet{HM14}. The resonant coupling causes some
of the mixed modes to have a high amplitude in the envelope and these are
called p-dominated mixed modes.  Theoretically, the oscillation modes
in red giants can be characterised by the dimensionless mode inertia,
$E$, which is a measure of the energy in the mode \citep[see,
e.g.,][]{Aerts10}.  The non-radial modes, in general, have much higher
inertia compared to the radial ones, except for the few p-dominated
mixed modes in each radial order which possess inertia of the same
order of magnitude as the radial modes. This is shown in
Fig.~\ref{fig:inertia} for a $1\msun$ red-giant model at an age of
12.5\,Gyr in the shell hydrogen-burning phase (Model~2 in
Fig.~\ref{fig:hrd}). The details of the model can be found in
Section~\ref{sec:appl_mod}. For the sake of clarity, we show only the
radial ($l=0$) and dipole ($l=1$) modes in this figure; the higher
degree non-radial modes ($l>1$) will have similar behaviour as the
dipole modes, though the exact distribution of mode inertia depends on
the evolutionary state of the star. The modes with the highest inertia
in each radial order are the g-dominated ones, with very little
amplitude in the p-mode cavity.  Observationally, only the modes with
low inertia attain significant heights in the power spectrum and can
be detected \citep{Dupret09}.  
\myfigure%
{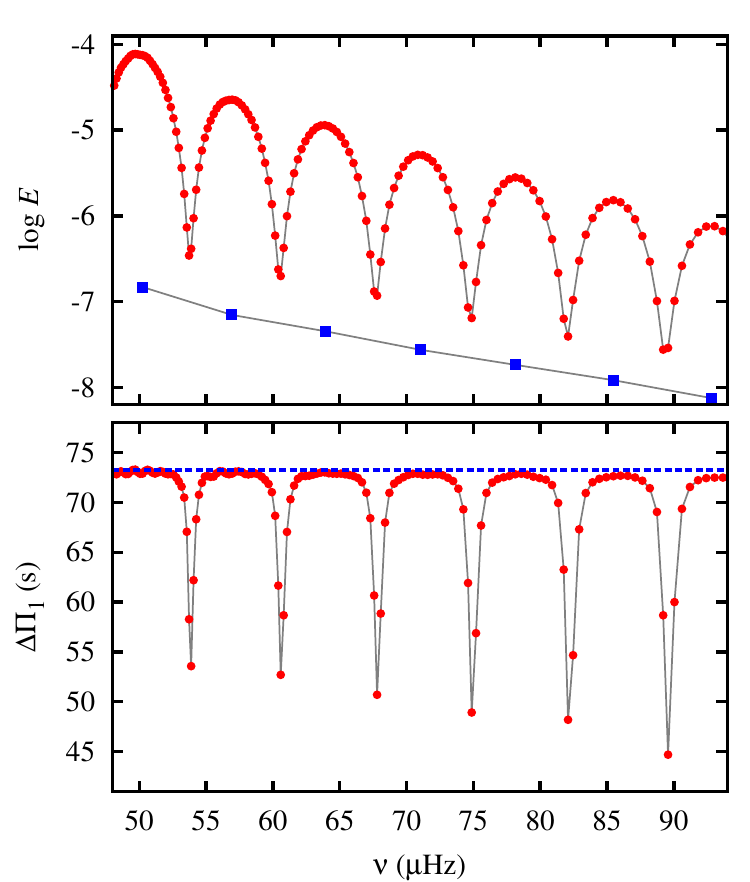}%
{%
Mode inertia and period spacing of a $1\msun$ red-giant model at an
age of 12.5\,Gyr (Model~2 in Fig.~\ref{fig:hrd}) are plotted as
functions of frequency for $l=0$ modes (blue squares) and  $l=1$ modes
(red dots). The top panel shows the mode inertia, while the bottom
panel shows the period spacing. The horizontal dashed blue line
indicates the asymptotic value of the dipole modes, \dpg\
($=73.48$\,s), as determined from Eq.~(\ref{eq:spacing}) using the
model.
}%
{fig:inertia}%
{}%

The g-modes are approximately equi-spaced in period (denoted by $\Pi$) with an asymptotic
value of the period spacing, \dplg, being given by 
\begin{equation}
\dplg =  \dfrac{\Pi_0}{\sqrt{l(l+1)}}(n_g+\epsilon_g)\,,
\label{eq:spacing}
\end{equation}
where
\begin{equation}
\Pi_0 = 2 \pi^2 \left( 
\displaystyle\ \int_\mathrm{g} \dfrac{N}{r} \mathrm{d}r  \right)^{-1} ,
\label{eq:brunt_int}
\end{equation}
$N$ being the Brunt-V\"{a}is\"{a}l\"{a} frequency. The integral is
carried out over the g-mode cavity.  This equal spacing in period
holds for the underlying pure g-modes in the core, as referred to as $\gamma$ modes by \citet{Aizenman77,Bedding14}. In reality, this
period spacing would be found in the most g-dominated modes (at the
inertia maxima in the top panel of Fig.~\ref{fig:inertia}), while the
modes with p-dominated character will have a smaller spacing in
period. This is illustrated in the bottom panel of
Fig.~\ref{fig:inertia}, where the dips in the period spacing
correspond to the minima in the inertia, while the g-dominated modes
maintain a period spacing very close to the asymptotic value given in
Eq.~(\ref{eq:spacing}).

In Fig.~\ref{fig:echelle}(a), the frequencies with azimuthal order $m=0$ of a $1\msun$ red giant
model at an age of 12.6\,Gyr (Model~3 in Fig.~\ref{fig:hrd}) are
plotted as a function of the period spacing $\Delta \Pi_1$. The
vertical ridge in this figure represents the value of the g-mode
period spacing, \dpg, while the p-dominated modes have much smaller
values of $\Delta\Pi_1$.  We do not expect a \dpg\ value higher than
the asymptotic one, and the only case where one can find such a value
in the observations is when one or more intermediate non-radial modes
have not been detected between two observed peaks in the power
spectrum.

If the frequencies of the $l=1$ modes are plotted as a function of $
\Pi_1$ modulo \dpg\ ($\Pi_1$ being the period of $l=1$ modes), one can
expect a vertical alignment of the $l=1$ modes with the same azimuthal order \citep[see,
e.g.,][]{Bedding11}. Such a diagram, known as the \ped, is shown in
Fig.~\ref{fig:echelle}(b) for the same stellar model as in
Fig.~\ref{fig:echelle}(a). The almost vertically aligned modes in this
diagram are the g-dominated dipole modes. The modes placed away from
this vertical ridge are more mixed in character, with the most
extremely deviant ones being the most p-dominated modes. At low
frequencies the ridge becomes curved as a result of the difference between the period spacing obtained asymptotically and from individual frequencies in the observable frequency range.

The g-mode period spacing of red giants can be used to distinguish
between their evolutionary phases, i.e., shell hydrogen-burning and
core helium-burning, as shown by \citet{Bedding11}. In their work the
vertical stacking in the \ped\ was shown to give the g-mode period
spacing, although no quantitative formulation to determine \dpg\ was
given. In the present work we describe an automated technique to
determine the g-mode period spacing by recognising the vertical
stacking of the g-dominated modes and the roughly symmetrical
deviation of the p-dominated modes from this vertical ridge in the
\ped.

The asymptotic value of the g-mode period spacing, obtained from
Eq.~(\ref{eq:spacing}), is usually slightly higher than the actual
value obtained from the frequencies of the model. For the particular
model shown in Fig.~\ref{fig:echelle}, for example, the asymptotic
value of \dpg\ is 62.15\,s, while the value obtained from the
frequencies is 62.05\,s. This slight difference is a reflection of the
validity of the asymptotic treatment. In any case, this difference is
much smaller than the uncertainty in \dpg\ that would be introduced by
the random uncertainties in the measurement of the stellar
frequencies. Thus, an estimate of \dpg\ from observed frequencies can
be considered as a reliable approximation to its asymptotic form as
given by Eq.~(\ref{eq:spacing}).
\myfigure%
{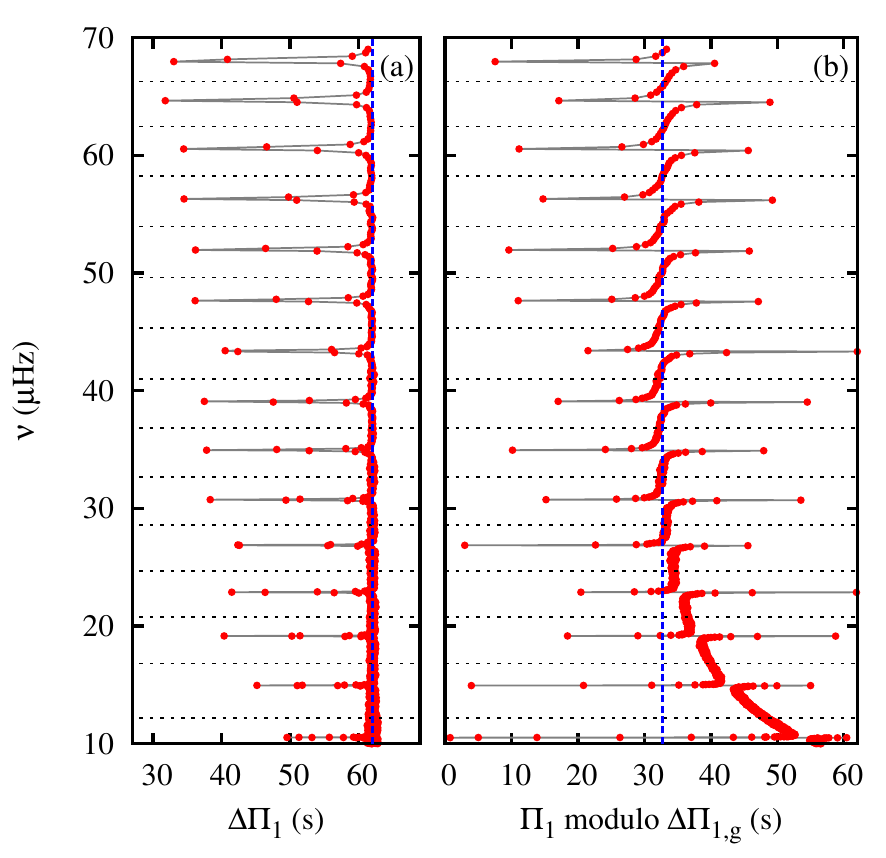}%
{%
(a) Frequency vs period spacing for $l=1$ modes for a $1\msun$
red-giant model at an age of 12.6\,Gyr (Model~3 in
Fig.~\ref{fig:hrd}).  The blue dashed line shows the g-mode period
spacing of high order g-modes.  (b) Period echelle diagram for the
$l=1$ modes of the model with $\dpg = 62.05$\,s. The blue dashed line
indicates the vertical alignment of the high-order g-dominated modes.
In both panels the red points connected with the light grey lines are
$l=1$ modes and the horizontal dotted lines represent the frequencies
of the radial modes.%
}%
{fig:echelle}%
{}%

\section{G-mode Period Spacing finder (GPS)}
\label{sec:tgps}

The \ped\ for red giants, using only oscillations with $m=0$ as shown in Fig.~\ref{fig:echelle}(b), has two
important features, namely, vertical alignment of the g-dominated
modes and the roughly symmetrical deviation of the p-dominated modes
from this vertical ridge. The inertias of the g-dominated modes are
much higher and their amplitudes and heights in the observed power
spectrum are correspondingly lower \citep{Dupret09}. This means that
only modes with lower inertias, i.e., modes with sufficient amplitude in the p-mode region can be
observed in the actual data even with the best available instruments
like \kep.  The period spacing of these observed p-modes (\dpl) is
much smaller than those of the g-dominated modes (\dplg).  However, it
is \dplg\ which has diagnostic power to probe the innermost layers of
the star through the asymptotic relation discussed above. Therefore, a
method to determine \dplg\ from \dpl\ is of considerable interest. For
$l=2$ and $l=3$, only the most p-dominated modes are usually visible,
if at all, which makes it difficult to draw any inference about the
period spacing. For $l=1$, however, several mixed modes can be
detected around the most p-dominated modes. Even though the period
spacing of these modes is far different from the period spacing of the
g-dominated modes, they contain information about the deep interior
through the coupling between the outer and inner oscillation cavities.
It turns out that it is possible to exploit the underlying vertical
alignment pattern of the \ped\ to estimate \dpg\ reliably from these
p-dominated mixed $l=1$ modes.  Indeed, the g-mode period spacing for
a few red giants was determined by \citet{Bedding11} by demanding such
a vertical alignment in the \ped.  However, they do not provide any
standard prescription to determine \dpg\ from an observed set of
frequencies.

\citet{Mosser12} further developed this using an empirical approach to
find p- and g-dominated mixed dipole modes combined with an asymptotic
development to obtain the asymptotic period spacing based on work by
\citet{Shibahashi79} and \citet{Unno89}. This development is based  on
the fact that eigenfrequencies are derived from an implicit equation
relating the coupling of the p and g waves:
\begin{equation}
\tan \theta_\mathrm{p} = q \tan \theta_\mathrm{g}\,,
\label{eq:taneq}
\end{equation}
where $\theta_\mathrm{p}$ and $\theta_\mathrm{g}$ are the p- and
g-wave phases. The dimensionless coefficient $q$ indicates the level
of mixing \citep{Mosser12}. This was then used to derive an expression
for dipole mixed modes coupled to a pure pressure dipole mode
($\nu_{n_\mathrm{p},l=1}$):
\begin{equation}
\nu=\nu_{n_\mathrm{p},l=1}+\frac{\Delta\nu}{\pi} \arctan \left( q \tan \pi
((\Delta\Pi_1\nu)^{-1} - \epsilon_\mathrm{g}) \right)\,,
\label{eq:tanfunc}
\end{equation}
where $\epsilon_\mathrm{g}$ is a phase term \citep{Mosser12}.
Eq.~(\ref{eq:tanfunc}) is essentially the mathematical expression of
the tangential shape of the mixed dipole modes in each p-mode order,
in which the coupling $q$ is connected to the steepness of the
``S''-shape and $\epsilon_\mathrm{g}$ represents an offset.

In the work presented here, we construct an algorithm to find \dpg\ in
a systematic search method that achieves the vertical stacked pattern
in the period-echelle diagram as indicated by \citet{Bedding11}. This
algorithm is called `GPS'. GPS has been tested on both stellar models
as well as actual observed frequencies of red-giant stars observed
with the \kep\ satellite. The algorithm is described in detail in the
next subsection. Although we do not use the function derived by
\citet{Mosser12} (Eq.~(\ref{eq:tanfunc})) in GPS, we have performed
tests to verify our results with such fits. In all cases we obtain
consistent results.

\subsection{Algorithm}
\label{sec:tgps_algo}

As can be seen from Fig.~\ref{fig:echelle}, the $l=0$ and $l=1$
frequencies of a red giant, if arranged in increasing order of
frequencies, appear as a dense spectrum of the dipole modes with a few
radial modes separating them at regular intervals.  This feature is
used by GPS to separate the entire range of $l=1$ modes into sets
referred to as ``bands''. A band is a set of $l=1$ modes lying between
two consecutive $l=0$ modes, i.e., one p-mode order. The horizontal dashed lines in
Fig.~\ref{fig:echelle} are the boundaries of the bands. For observed
spectra, of course, the number of dipole modes in a band is much less
than that for a model.

The g-dominated dipole modes of high order are equally spaced in
period.  Thus for a correct choice of the period spacing \dpg\ to
construct the \ped, these g-dominated $l=1$ modes in a band would
align to form a nearly vertical central ridge. Since each band is
bordered by two radial modes, the presence of an underlying $l=1$
p-mode is expected around the middle of the band. This mode would give
rise to a few p-dominated mixed modes which would be farthest from the
central vertical ridge of the band. In moving from one radial mode
to the next, a horizontal ``wrap around'' is thus expected in the
\ped. The presence of such a shift from the extreme right to the
extreme left is considered as a necessary condition for a band to be
considered for vertical alignment. 
	
Usually the mixed p-dominated modes are expected in a short frequency
range centred around the underlying $l=1$ p-mode.  This would imply
that these modes would be approximately at equal distances from the
central ridge in the \ped, creating symmetrical upper and lower tails
of the ridge, i.e., this symmetry reflects an implicit assumption of $\epsilon_g \sim 0.5$.  However, this symmetry is not perfect as the
frequencies of the mixed modes would depend on the exact frequencies
of the underlying pure p- and g-modes as well as the extent of mixing
on either side of the central mode.

Therefore, ideally at the correct value of \dpg\ in the \ped, all the
bands should align vertically to form the central ridge with the mixed
p-dominated modes forming roughly symmetrical tails on either side.
Since the g-dominated modes also have mixed character, the central
ridge in the bands are not perfectly vertical but form a ``S-shaped''
structure with long tails as seen in Fig.~\ref{fig:echelle}.  The aim
is to obtain this alignment pattern in the \ped\ by scanning through
possible values of \dpg\ and choosing the correct value which achieves
such a pattern.

The algorithm implemented by GPS consists of the following steps to
arrive at the correct g-mode period spacing value.  The abscissa of
the \ped\ is henceforth denoted by $\Pi_\mathrm{r} \equiv \Pi_1 \mod
\dpg \equiv \Pi_1 \% \dpg$.

\begin{enumerate}
\item 
\textit{Lower frequency threshold}: The asymptotic period spacing given
by Eq.~(\ref{eq:spacing}) is valid only for high order g-modes. At low
frequencies, the modes have even higher orders compared to the modes that are in the observable frequency range and hence would give a period spacing based on individual frequencies closer to the asymptotic value. As only modes in the observable frequency range are considered here, the difference between the period spacing obtained from individual frequencies and the asymptotic value is manifested as a curvature in the \ped, as seen in Fig.~\ref{fig:echelle}.  To account for this a lower
threshold on frequency is necessary before GPS can be applied. However,
it is difficult to arrive at a specific threshold frequency in an
analytic fashion. Different criteria for the lower threshold were tried
for a large number of models spanning the entire red-giant branch
(RGB) and the following prescription was adopted heuristically:
\begin{equation}
\nu_{\rm{threshold}} = 
\begin{cases}
7.5\,\muHz, &\text{for}\ \nu_{\rm{max}} < 15\,\muHz \\
0.5\,\nu_{\rm{max}}, &\text{for}\ 15\,\muHz \leq \nu_{\rm{max}} \leq 160\,\muHz \\
80\,\muHz, &\text{for}\ \nu_{\rm{max}} > 160\,\muHz 
\end{cases}
\label{eq:cutoff}
\end{equation}
Successful tests with several observed stars confirmed this choice.

\item 
\textit{Band Identification}: In the \ped, a band refers to the $l=1$
modes (points in  Fig.~\ref{fig:echelle}) lying between two successive
$l=0$ modes (dotted lines in the same figure).  The bands are indexed
starting from the lowest frequencies to the highest ones. For a band
to be considered for the vertical alignment, it has to satisfy two
conditions. First, the \pir\ for the highest frequency of a band must
be less than that for the lowest frequency of the band immediately
above, i.e., the \pir\ value increases from the top end of one band to
the bottom end of the next band. Second, one and only one nearly
horizontal shift between the two most p-dominated modes in a band,
i.e., from the extreme right to the extreme left of the \ped\ should
be present. Any band which does not satisfy these two conditions is
not used in the vertical alignment in the \ped. GPS proceeds to the
next step only if the number of bands which do not satisfy these
conditions is less than or equal to two.   
\myfigure%
{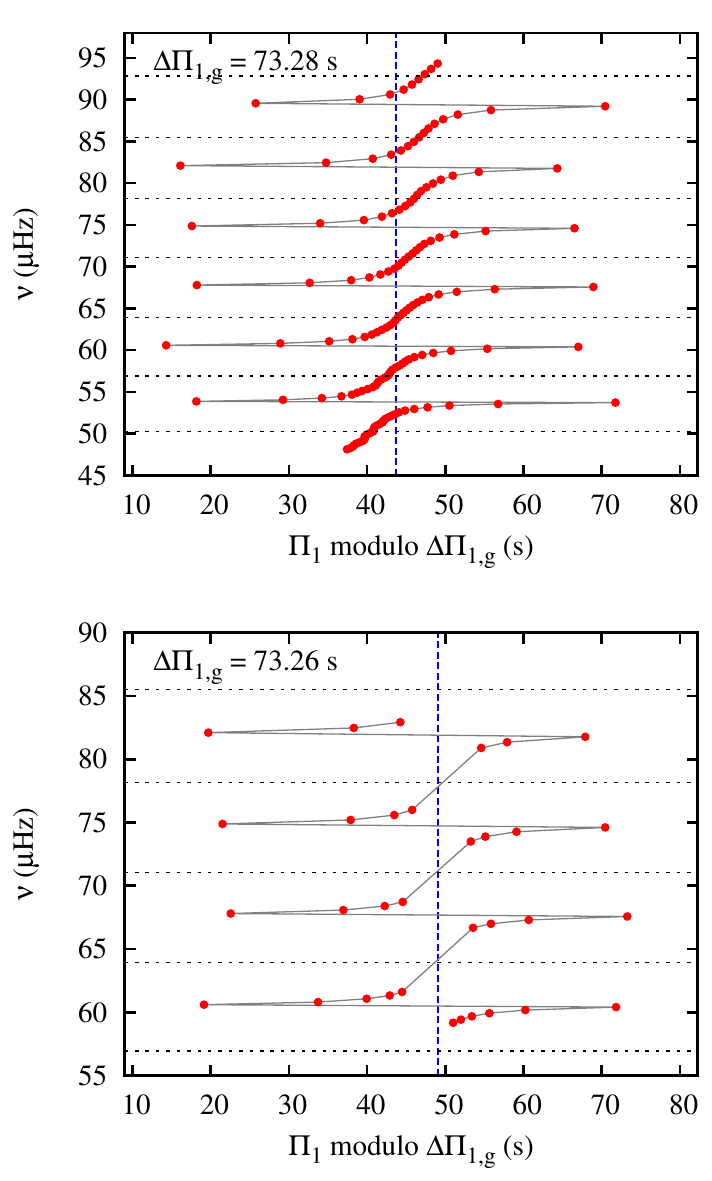}%
{%
Two \ped s with different density of $l=1$ modes for a $1\msun$
red-giant model are shown. The symbols and the lines have the same
meaning as in Fig.~\ref{fig:echelle}. The top panel shows the \ped\
for all the $l=1$ modes in a frequency range spanning 7 radial
orders, with $\dpg = 73.28$\,s as determined by GPS. The lower panel
shows the \ped\ for only a subset of the most p-dominated $l=1$ modes
in a smaller frequency range consisting of 5 radial orders with $\dpg
= 73.26$\,s. To illustrate the symmetrical distribution of the most
p-dominated modes, the abscissae have been offset slightly in both
diagrams.
}%
{fig:GPS_echelle}%
{}%

\item 
\textit{Vertical Alignment}: This step quantifies the vertical
stacking of the central g-dominated $l=1$ modes in the \ped\ for a
given trial value of \dpg.  This is done by first estimating the
position of the most g-dominated mode at the upper boundary of each
band in the \ped\ and then calculating the dispersion of these
positions for all bands from an average vertical ridge.  This
dispersion would later be minimised to obtain the correct value of
\dpg.

The location of the most g-dominated mode at the upper boundary of
each band is estimated as follows.  At the upper boundary of a band
(say, the $i^{\rm{th}}$ band), the weighted average value of \pir\ of
the highest dipole mode in that band (denoted by the index
$(i,\mathrm{h})$) and the lowest dipole mode in the next band (denoted
by the index $(i+1,\mathrm{l})$) is calculated.  These two $l=1$ modes
are the immediate neighbours of the radial mode separating the
$i^{\rm{th}}$ and the $(i+1)^{\rm{th}}$ bands (denoted by
$\nu_{l=0}^{(i,i+1)}$). For observed frequencies, 
the differences in frequency between each of these $l=1$ modes 
and the radial mode between them can be significant and may not even 
have similar values. To account for this 
difference in frequency we incorporate a weighing.
The weights are taken as the differences in frequency between the
respective $l=1$ mode and the neighbouring $l=0$ mode.  The weighted
mean thus calculated is an estimate of the \pir\ value of the
(possibly unseen) most g-dominated mode at the boundary of the two
bands. It is denoted by $\pirgi$ for the $i^{\rm{th}}$ band and  is
calculated as follows:
\begin{equation}
\pirgi = \dfrac{y^{(i)}\pir^{(i+1,\mathrm{l})} + y^{(i+1)}\pir^{(i,\mathrm{h})}}{y^{(i+1)} + y^{(i)}}  
\label{eq:pirgi}
\end{equation}
where
$y^{(i+1)} = \nu_{l=1}^{(i+1,\mathrm{l})} - \nu_{l=0}^{(i,i+1)} $ 
and 
$y^{(i)} = \nu_{l=0}^{(i,i+1)} - \nu_{l=1}^{(i,\mathrm{h})}$.

From all the values of \pirgi\ calculated for each accepted band,
their arithmetic mean, \pirg, is determined which is called the
average midpoint, for a particular choice of \dpg. This is considered
as the mean position of the central vertical ridge. The squared
deviation \chisqvert\ of the $\pirgi$ values for each accepted band
from this mean position is calculated as 
\begin{equation}
\chisqvert =  \sum_{i=1}^N \dfrac{(\pirgi - \pirg)^2}{(\delta\pirgi)^2} 
\end{equation}
where $\delta \pirgi = \dfrac{1}{2} \sqrt{%
(\delta\pir^{(i+1,\mathrm{l})})^2 + (\delta\pir^{(i,\mathrm{h})})^2 }$
is the uncertainty in \pirgi\ and the total number of accepted bands
is $N$.  Here,  
 $\delta\pir^{(i+1,\mathrm{l})} = %
 \delta \nu_{l=1}^{(i+1,\mathrm{l})} /%
 (\nu_{l=1}^{(i+1,\mathrm{l})})^2 $
and
  $\delta\pir^{(i,\mathrm{h})} = %
 \delta \nu_{l=1}^{(i,\mathrm{h})} /%
 (\nu_{l=1}^{(i,\mathrm{h})})^2 $,
where $\delta \nu$ refers to the uncertainty in the observed
frequencies.  We have not considered the uncertainties in the weighing
factors $y^{(i)}$ and $y^{(i+1)}$.

Among all trial values of \dpg, higher priorities are given to the cases with
lesser number of unacceptable bands. Minimisation of \chisqvert\ alone among
the cases with lowest number of unacceptable bands would give the best
vertical alignment of the bands. In Fig.~\ref{fig:GPS_echelle}, in both
period echelle diagrams, there are no unacceptable bands and a vertical
alignment has been achieved.

\item 
\textit{Symmetrical distribution}: As described before, besides the
vertical alignment of the g-dominated modes to form the central ridge,
a nearly symmetrical distribution of the p-dominated modes is also
expected around the central ridge in the \ped, i.e., $\epsilon_g \sim 0.5$. This symmetrical
positioning of the tails on either side can be seen in both the period
echelle diagrams in Fig.~\ref{fig:GPS_echelle}. This criterion is
implemented in GPS in the following way.  The arithmetic mean of the
two extreme p-dominated mode frequencies on either side of the central
ridge in the $i^{\rm{th}}$ band is
\begin{equation}
\pirpi = \dfrac{\pirir+\piril}{2},
\end{equation}
where \pirir\ and \piril\ refer to the $l=1$ mode frequencies located
on the extreme right (\nuir) and left (\nuil) of the period echelle
diagram in the $i^{\rm{th}}$ band respectively. These are the
frequencies between which the ``wrap-around'' mentioned earlier occurs
in the $i^\mathrm{th}$ band. For a correct choice of \dpg\ these modes
should be the most p-dominated modes in the $i^\mathrm{th}$ band,
lying on either side of the underlying pure $l=1$ p-mode. The \pir\
value of this underlying pure p-mode would be close to either 0 or
\dpg\ on the \ped, that is, it would be farthest removed from the pure
g-modes which constitute the central ridge. Therefore, the two most
p-dominated modes would be at the two extreme ends of the \ped\ and
the average of their \pir\ values, \pirpi, should be close to the
central ridge, \pirg. On the other hand, for a wrong choice of \dpg\
both these modes would be on the same side of the \ped\ and the value
of \pirpi\ would be far removed from \pirg; this would indicate an
asymmetrical distribution.

The squared deviation of the \pirpi\ values from the central vertical
ridge, \pirg\ (determined in the previous step) is calculated as
\begin{equation}
\chisqsymm =  \sum_{i=1}^N \dfrac{(\pirpi - \pirg)^2}{(\delta\pirpi)^2} 
\end{equation}
where $\delta\pirpi = \dfrac{1}{2} \sqrt{ (\delta\pirir)^2 +
(\delta\piril)^2 }$. Here, $\delta\pirir =
\dfrac{\delta\nuir}{(\nuir)^2}$ and $\delta\piril =
\dfrac{\delta\nuil}{(\nuil)^2}$.

Minimising \chisqsymm\ alone in the same way as the minimisation of
\chisqvert\ gives the best possible symmetrical distribution around
the central ridge in the \ped\ which can be seen in
Fig.~\ref{fig:GPS_echelle}. 

\item 
\textit{Final minimisation}: To obtain the g-mode period spacing, we
minimise the total \chisqtot\ which has contributions from both
\textit{Vertical Alignment} and \textit{Symmetrical Distribution} in a
weighted manner: 
\begin{equation}
\chisqtot = (1-a)\chisqvert + a\chisqsymm
\end{equation}
where $a$ is a weighting factor reflecting the emphasis on the
\textit{Symmetrical distribution}. It was found that a fixed value of
$a$ is not suitable for all stars and it has to be varied over the RGB
to get a sensible estimate of \dpg.  The idea of symmetrical
distribution rests crucially on the assumption that there are two
p-dominated modes with very low inertia lying midway between two
radial modes. This turns out to be the case in the lower RGB, but not
necessarily higher on the RGB. It was found that in the upper RGB $a$
has to be very small. In the lower RGB $a$ is restricted to 0.5 so
that equal weightage is given to \textit{Vertical Alignment} and
\textit{Symmetrical Distribution}. Different values of $a$ were tried
for a large number of models spanning the entire RGB and the following
working relation for $a$ was found heuristically. 
\begin{alignat}{3}
a & =  0.0 &&~\rm{for}\ \Delta\nu < 1.65\,\muHz  \nonumber \\
\log a & = A\Delta\nu + B &&~\rm{for}\ 1.65\,\muHz \leq \Delta\nu \leq 12.50\,\muHz \\
a & =  0.5 &&~\rm{for}\ \Delta\nu > 12.50\,\muHz  \nonumber
\label{eq:weightage}
\end{alignat}
Here $A$ and $B$ are constants which can be determined to be $A =
0.1566\,\muHz^{-1}$ and $B = -2.2584 $ from the heuristic exercise.
Such a choice of $a$ was found to be appropriate for several observed
stars as well. 

As mentioned earlier, highest priority is given to the cases with
least number of unacceptable bands. Thus the final value of \dpg\ is
the one which minimises \chisqtot\ among all the cases with least
number of unacceptable bands.  

\end{enumerate}

In the remainder of the paper we show the application of GPS on
red-giant models, as well as observed stars, to successfully determine
their g-mode period spacing \dpg\ along with estimates of
uncertainties in the results.

\section{Application of GPS}
\label{sec:appl}

\subsection{Finding \dpg\ of models with GPS}
\label{sec:appl_mod}

\myfigure%
{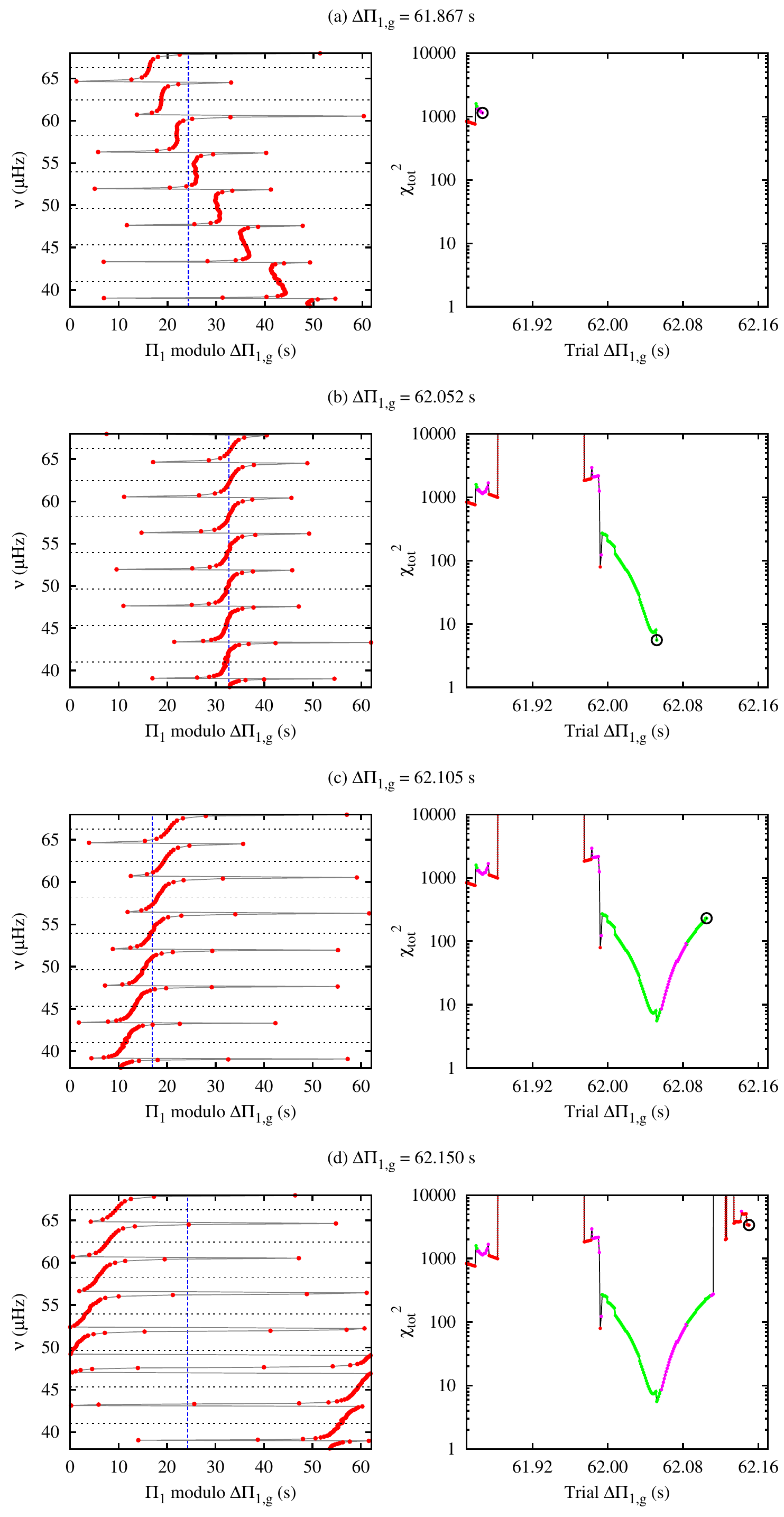}%
{%
Four snapshots illustrating the method of GPS are shown. Trial values
of \dpg\ are in increasing order from (a) to (d). In each part, the
left panel shows the period echelle diagram, where symbols have the
same meaning as in Fig.~\ref{fig:echelle}. The right panel shows
\chisqtot\ as a function of the trial value of \dpg. The trial value
used in the left panel corresponds to the last (encircled) point in
the corresponding right panel and is indicated at the top of each pair
of panels. The colours of the points in the right panel indicate the
number of unacceptable bands (red for 2 or more, magenta for 1 and
green for 0).
}%
{fig:GPS_movie}%
{}%

Since the full model frequency spectrum includes the g-dominated
modes, the determination of \dpg\ is much more straightforward than in
observed spectra. In fact, these modes influence the vertical
alignment procedure very strongly, while the p-dominated modes play a
dominant role in the symmetrical distribution. To illustrate the
procedure of GPS a 1\,\msun\ red giant is considered using all the
calculated eigenfrequencies. 

All the models considered in this work have been constructed using the
MESA stellar evolution code \citep[][version 4798]{Paxton11}. The
models used standard physics such as the OPAL equation of state
\citep{Rogers02}, OPAL high temperature opacities \citep{Iglesias96}
supplemented by the low temperature opacities from \citet{Ferguson05}.
The nuclear reaction rates were from NACRE \citep{Angulo99} for all
reactions except $^{14}$N($p$,$\gamma$)$^{15}$O and
$^{12}$C($\alpha$,$\gamma$)$^{16}$O, for which updated rates of
reaction from \citet{Imbriani05} and \citet{Kunz02} were used.
Convection was modelled using the standard mixing length theory
\citep{Cox68}. Diffusion of helium and heavy elements was not included
in the models. We used the standard solar mixture of \citet{GS98}.
Adiabatic pulsation frequencies of the models were computed with the
ADIPLS code \citep{JCD08}.

Fig.~\ref{fig:GPS_movie} shows the step-by-step procedure followed by
GPS. In each of the four parts, the \ped\ along with the total \chisq\
(\chisqtot) is plotted as \dpg\ is varied continuously through trial
values.  In part (a) of the figure, the trial \dpg\ is less than the
correct value. Although a number of bands are symmetrically
distributed, they are not vertically aligned. In (b) the trial \dpg\
is very close to the correct value so that vertical alignment and
symmetrical distribution are simultaneously achieved. The trial \dpg\
in (c) is slightly greater than the correct value so that the bands
are nearly vertically aligned but not symmetrically distributed.  In
part (d) the trial \dpg\ is much greater than the correct value, and
thus the bands are neither vertically aligned nor symmetrically
distributed. Further, at this value one of the bands is unacceptable
because there are more than one shift from the extreme right to the
extreme left. The departure of the trial value from the correct value
is reflected both in the high \chisqtot\ value and the indication of
the number of unacceptable bands.  The best symmetrical vertical
alignment is obtained when \chisqtot\ passes through a global minimum,
as shown in Fig.~\ref{fig:GPS_movie}. The \dpg\ for this model is
found to be 62.052\,s by GPS, which agrees quite well with the
asymptotic value of 62.149\,s calculated using Eq.~(\ref{eq:spacing}).

\myfigure%
{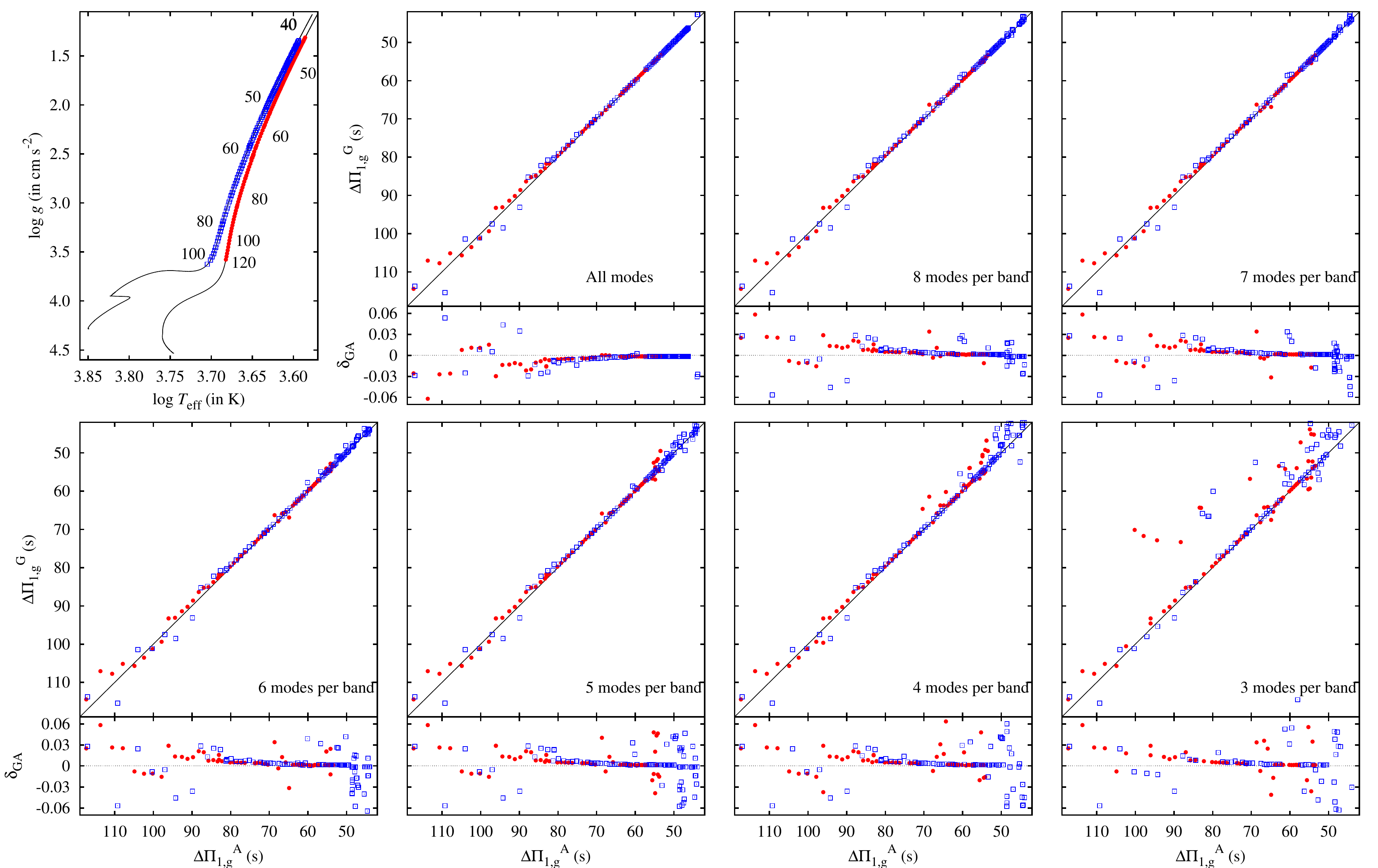}%
{%
Comparison of \dpg\ values obtained with GPS for models with the
corresponding asymptotic values are shown for different choices of
number of modes in each band. The top left panel shows the tested
models of 1\,\msun (red filled circles) and 1.5\,\msun (blue empty
squares) in a \logtg\ diagram. The asymptotic values of \dpg\ (in s)
from the models are indicated alongside the evolutionary tracks (black
continuous curves). In the upper part of each of the other panels the
\dpg\ values obtained from GPS ($\dpg^\mathrm{G}$) are plotted against
the asymptotic values ($\dpg^\mathrm{A}$). In the lower part the
fractional differences between the two, $\delta_\mathrm{GA} \equiv 1 -
\dpg^\mathrm{G}/\dpg^\mathrm{A}$, are shown. 
}%
{fig:asymp_comp_modes}%
{*}%

We applied GPS to a series of evolutionary models of masses 1\,\msun\
and 1.5\,\msun\ on the red-giant branch. The comparison of the values
of \dpg\ with the respective asymptotic values from these models are
shown in Figs.~\ref{fig:asymp_comp_modes}
and~\ref{fig:asymp_comp_bands}.  We tested the effect of both the
total number of available bands and the number of modes available in
each band on the results produced by GPS. 

In the first test we used 5 bands around the frequency of maximum
oscillation power, \numax, computed by the scaling relation of
\citet{KB95} from the model, but varied the number of $l=1$ modes in
each band from 3 to 8. These modes were the ones with lowest inertia
values in each band. We also tested GPS using all the available modes
in 5 bands. The results of the comparison of the values obtained from
GPS with the asymptotic values from the models are shown in
Fig.~\ref{fig:asymp_comp_modes}. In general, the \dpg\ values obtained
by GPS agree very closely with the asymptotic values. Although the
difference between the two increases as the number of modes in each
band are decreased, for at least 5 modes per band, the \dpg\ value found
by GPS is within 5\% of the asymptotic value for almost all the models
that we considered. Even with only 4 or 3 modes per band, GPS can find
the correct value within this limit for models with \dpg\ values
between 50 and 110\,s.
\myfigure%
{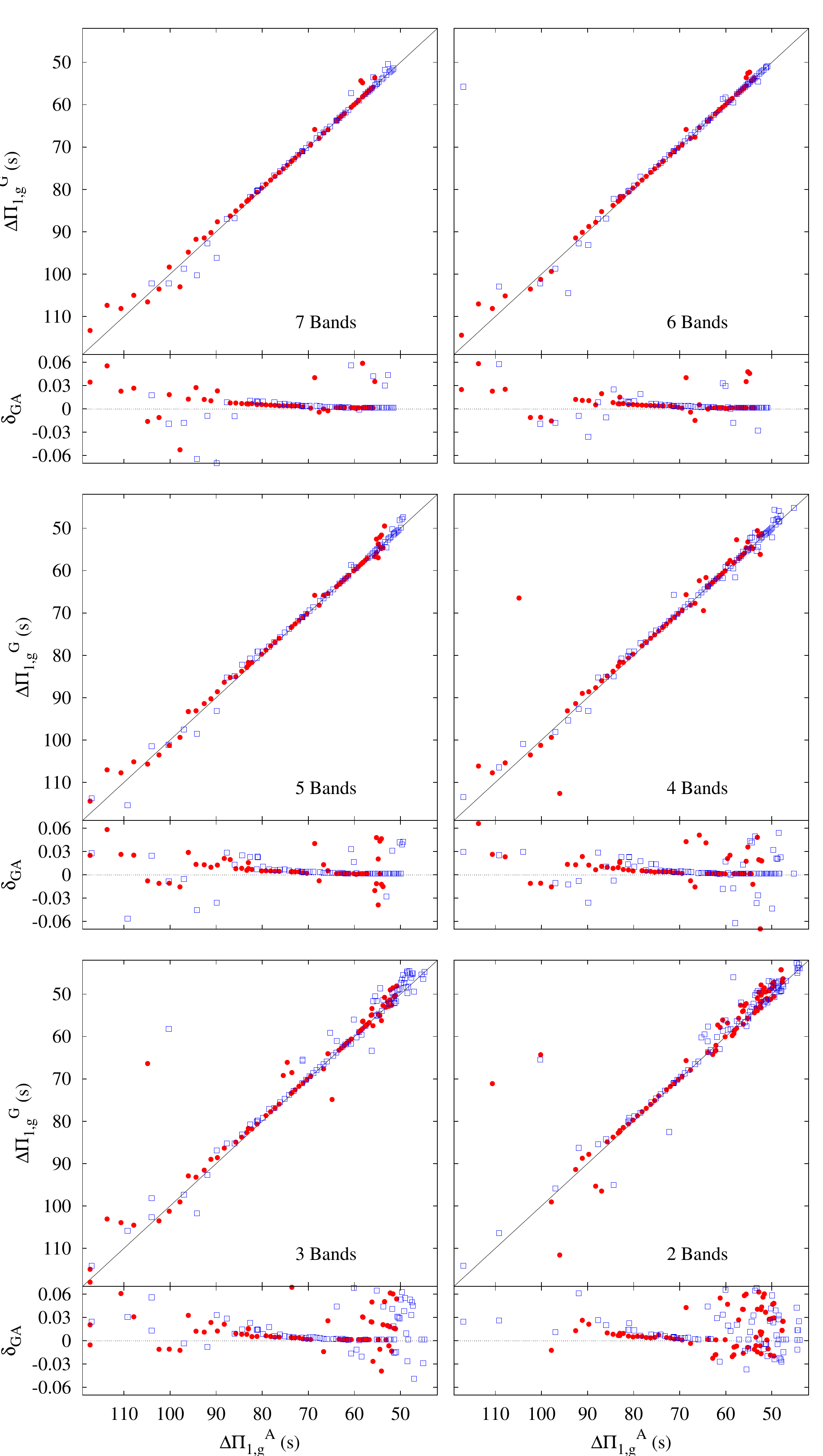}%
{%
Comparison of \dpg\ values obtained with GPS for models with the
corresponding asymptotic values are shown for different choices of
total number of bands. The symbols have the same meaning as in
Fig.~\ref{fig:asymp_comp_modes}. 
}%
{fig:asymp_comp_bands}%
{}%

In the second test we used only five $l=1$ modes with the lowest
inertia values in each band but varied the total number of bands from
2 to 7. As expected, the \dpg\ values are determined with higher
accuracy when using a larger number of bands. The comparison of these
results with the corresponding asymptotic values of \dpg\ are shown in
Fig.~\ref{fig:asymp_comp_bands}.  Again, the values from GPS agree
with the asymptotic values to within 5\% for nearly all cases with at
least 4 bands. Using only 2 or 3 bands we can still get values within
this limit for models with \dpg\ greater than 60\,s. 

In general, we find that GPS can be reliably applied only to
frequencies of models with \logg\ values between 3.6 and 1.3 (see top
left panel of Fig.~\ref{fig:asymp_comp_modes}). We recommend that GPS
should be applied to observed RGB stars only within such limits of
\logg.

\subsection{Application of GPS on observed frequencies}
\label{sec:appl_obs}

For the observed frequencies, the number of modes available is much less
compared to models. Specifically, the g-dominated modes which
constitute the central ridge of the vertical pattern are absent and
only the p-dominated modes are present in the observed spectrum.
However, provided that a few mixed dipole modes are present in the
spectrum, GPS can successfully determine \dpg. Of course, the
uncertainty of the estimated value reduces with the number of modes
that are detected close to the vertical ridge. We elaborate on the
uncertainties in more detail in Section~\ref{sec:appl_unc}. To
illustrate the application of GPS to real stellar data, we show here
the results of three red-giant stars, \kiconezero, \kicnineone, and
\kicfiveeight, observed by the \kep\ space mission
\citep{Borucki08,Gilliland10} during its nominal operation.  The
position of these stars are shown on the HR diagram in
Fig.~\ref{fig:hrd}.
\myfigure%
{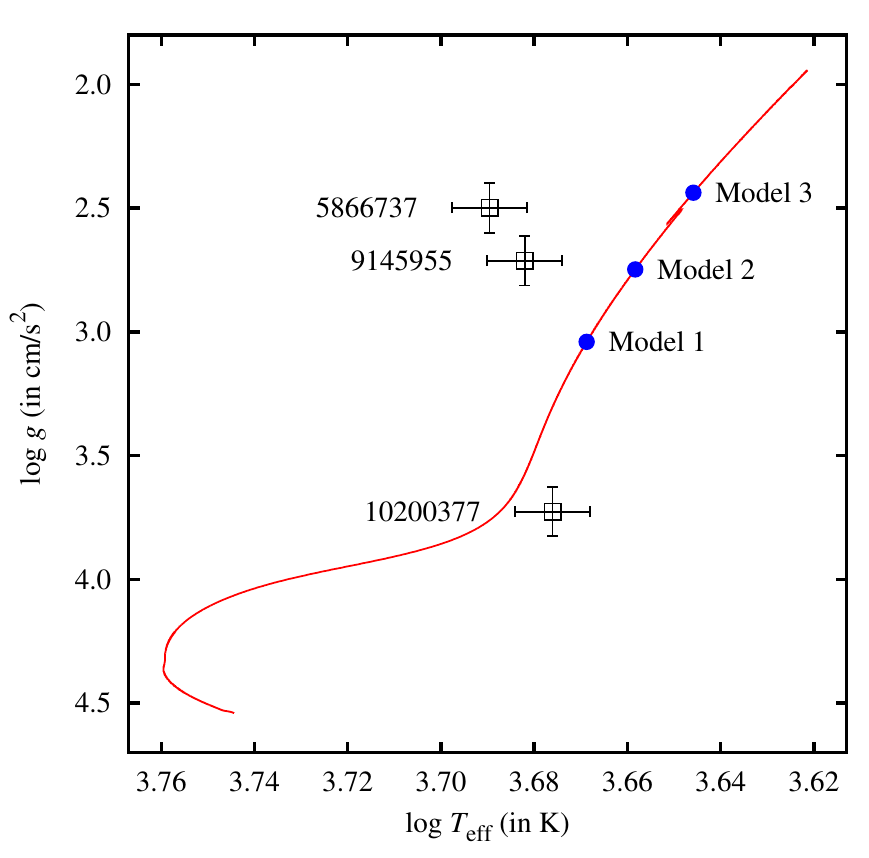}%
{%
The position of three red-giant stars observed by \kep\ are shown on
the \logtg\ diagram. The effective temperatures are taken from
the \kep\ Input Catalogue (KIC) for \kicfiveeight\ and from the APOGEE
catalogue for the other two stars. The \logg\ values are from KIC. Also
shown, by the red line, is the evolutionary track of a 1\msun\ model.
The blue dots indicate the three models that we discuss in this paper.
}%
{fig:hrd}%
{}%

The photometric timeseries were gathered during the first 12 quarters
of the \kep\ mission, where quarter 0 lasted for only 10 days, quarter
1 lasted for 30 days and all other quarters for nominally 90 days.
This provided us with timeseries data of over a 1000 days length. We
used observations with a 29.4-minute sampling. For more details about
\kep\ data and their treatment, we refer to e.g.
\citet{Jenkins10,Garcia11}.

Fig.~\ref{fig:echelle_obs} shows the period echelle diagrams for these
three red giants: \kiconezero\ with $\dpg=81.59$\,s, \kicnineone\ with
$\dpg=76.98$\,s, and \kicfiveeight\ with $\dpg=68.51$\,s,
respectively.  In this figure,  it can be seen from the left panels
that most of the $l=1$ modes in the observed spectrum have low values
of period spacing and thus are the p-dominated modes. The g-dominated
modes which have higher and uniform value of period spacing as seen in
Fig.~\ref{fig:echelle}(a) are absent in the observed spectrum.
Nevertheless, GPS is able to identify a value of \dpg\ which produces
the expected vertical alignment of the dipole modes in the \ped, as
seen in the right panels of Fig.~\ref{fig:echelle_obs}.  A few of the
observed modes have period spacing of nearly double the average value,
which indicates that a neighbouring dipole mode has not been detected.
Thus GPS can also help in flagging these missed modes in the power
spectrum. 
\myfigure%
{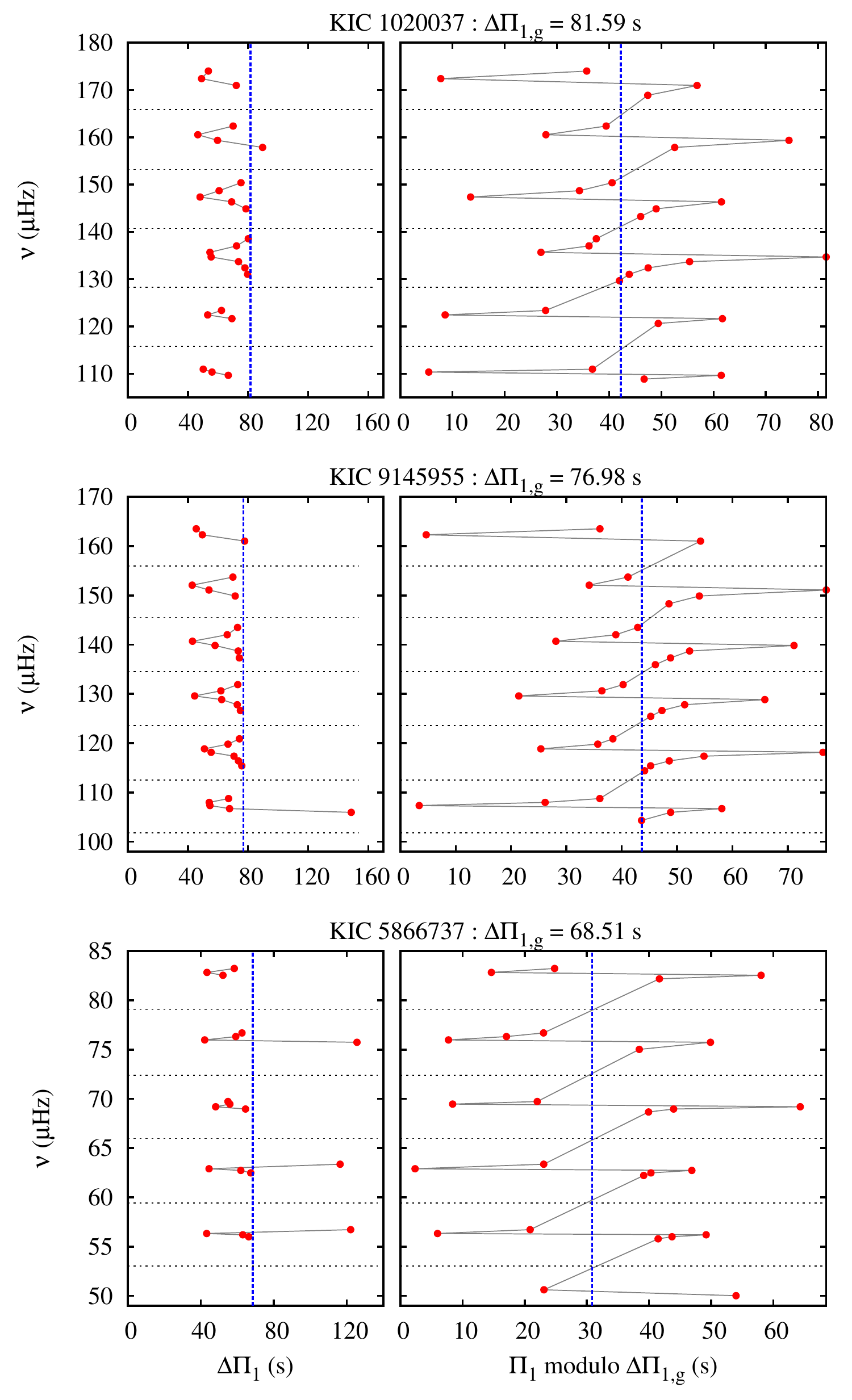}%
{%
Results obtained with GPS for three red giants with \kep\ frequencies:
\kiconezero\ (top) \kicnineone\ (middle), and \kicfiveeight\ (bottom).
The left panel in each plots the frequencies as a function of observed
$\Delta \Pi_1$, while the right panel shows the \ped.  The symbols and
the lines have the same meaning as in Fig.~\ref{fig:echelle}.  The
value of \dpg\ determined by GPS is indicated at the top of each pair
of panels.  In the left panels of the latter two stars the large
values of $\Delta\Pi_1$ for a few modes indicate that a neighbouring
$l=1$ mode has not been detected. 
}%
{fig:echelle_obs}%
{}%

\subsection{Estimation of the Uncertainties in \dpg}
\label{sec:appl_unc}

So far, we have described the application of GPS to either model
frequencies or to observed frequencies without regard to their
associated uncertainties.  In observed data the uncertainties in the
frequency values will affect the determination of \dpg\ through GPS.
Also it is important to know the maximum permissible uncertainty in
observed frequencies for which GPS can be applied reliably. 

The effects of the uncertainties in the frequencies are considered by
repeating the procedure carried out by GPS for 10000 realisations of
the data, produced by perturbation of the frequencies by random values
corresponding to a normal distribution with standard deviation equal
to the given 1$\sigma$ uncertainty in the frequencies. This approach is justified by the fact that only the frequencies of $l=1$ modes which are narrow due to the high inertia are perturbed.

It is conceivable that for low values of the assumed uncertainty in
the frequencies GPS would return values of \dpg\ which are close to
the actual value for a majority of the realisations. In the Monte
Carlo exercise, this would be reflected in a single peak in the
histogram of the \dpg\ values centred around the actual value. The
median value of \dpg\ for all the realisations would be the estimated
value and the width of this peak, measured in terms of 34\% area
coverage on either side of the median value, would be a fair
representation of the uncertainty in \dpg. 

\myfigure%
{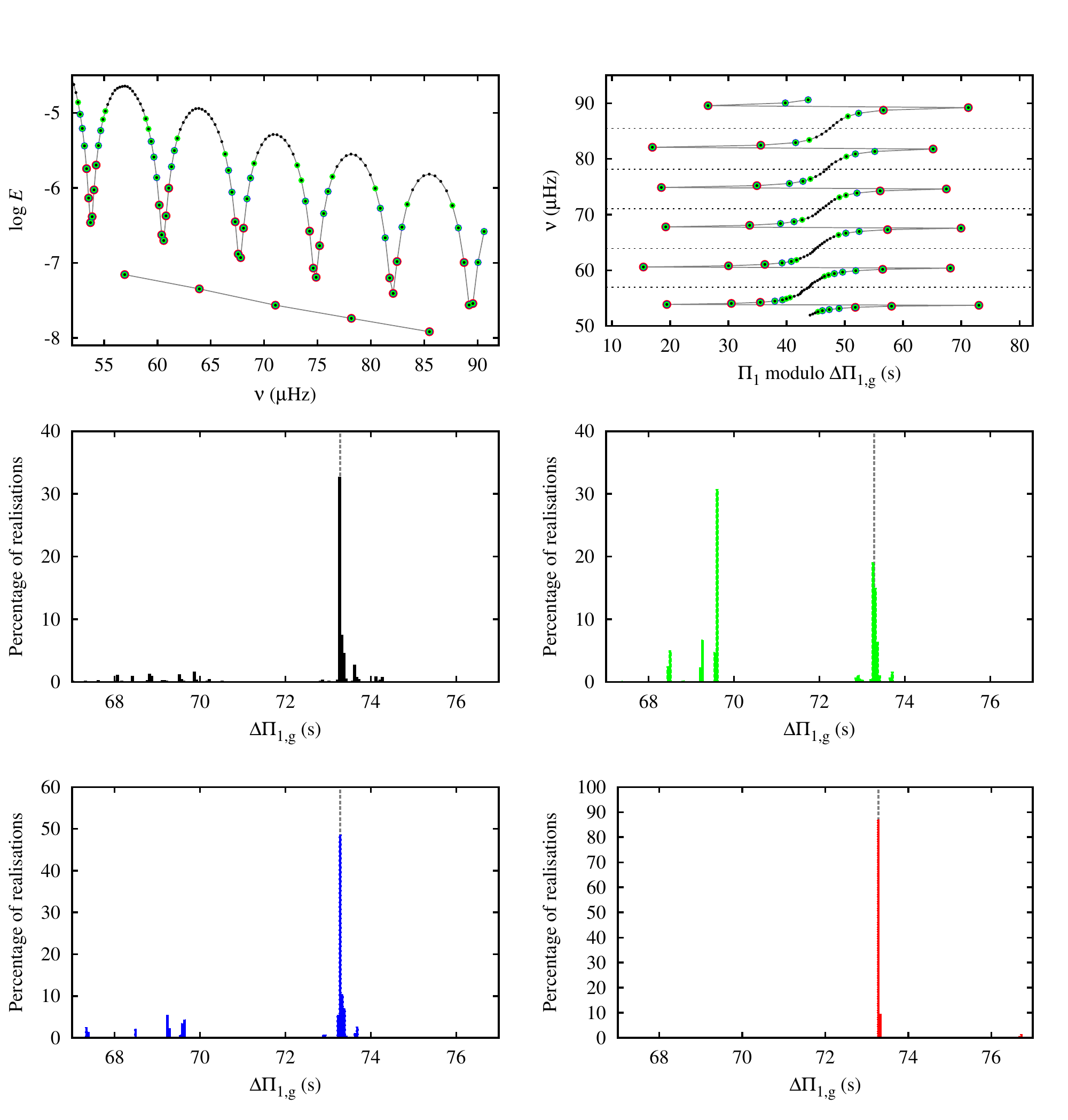}%
{%
Results of the Monte Carlo exercise with the frequencies of Model~2
with a uniform uncertainty of 0.01\muHz. The top left panel shows the
inertia of the $l=0$ and 1 modes of the model. The four different
choices of the mode sets used for the MC exercise are shown in
different colours.  Solid black dots represent the full set of
frequencies, while green, blue and red circles denote sets with
progressively lower cutoffs applied on the mode inertia. The top right
panel shows the \ped\ where \dpg\ has been set to the value of
73.28\,s found from the unperturbed model frequencies. The symbols and
lines are similar to that of Fig.~\ref{fig:echelle} with the different
colours corresponding to the four sets as depicted in the top left
panel.  The remaining four panels show the histograms of the \dpg\
values found in the MC exercise for the four sets, again corresponding
to the same colours as depicted in the top left panel. In each of the
bottom four panels the vertical dotted line shows the value of \dpg\
obtained for the unperturbed model frequencies.
}%
{fig:model2_inertia_cut}%
{*}%

As the uncertainty in the frequencies increases, however, other values
of \dpg\ become likely and we get distinct multiple peaks in the
histogram at discrete values away from the correct value of \dpg. In
such cases, the overall median and the uncertainty based on 34\% area
coverage on either side of that value would not be a true
representation of the situation.  Instead, we have chosen to report
the three most probable values and associated uncertainties along with
respective probabilities. Graphically, these would be the three most
significant peaks in the histogram with highest area coverage.
Specifically, we define the peaks in the histogram as follows. The bin
with the highest population is chosen along with its neighbouring bins
till the value in a bin drops to below 10\% of the maximum value.
These bins together constitute the most significant peak. From among
the remaining bins the one with the next highest value is chosen and
the same criterion is applied to include its neighbouring bins. These
bins together form the second highest peak, and so on.   For each peak
we provide the median value under that peak,  its $1\sigma$
uncertainty measured in the way described above, and the probability,
$p$, of that value, which is the fractional area under that peak.
 
This exercise was carried out with three red-giant models as well as
the above-mentioned three observed red giants. The models are chosen
such that they lie at different positions on the red giant branch and
are labelled as Model~1, Model~2 and Model~3, respectively, in
increasing order of their luminosities. All the models lie on the same
evolutionary track of a 1\,\msun\ star with solar-like chemical
abundances (see Fig.~\ref{fig:hrd}). 

Since for models the entire theoretically computed frequency spectrum
is available, we test GPS for a variety of cases with varying
selection of frequencies and the associated uncertainties. The
selection of frequencies was carried out in two ways.  First, we have
considered three different ranges of frequencies corresponding to 4, 5
and 6 times the large separation, each centred around \numax, i.e., we
chose the range of frequencies to be $\numax \pm k\Delta\nu$, for
$k=2.0,2.5,3.0$. Second, we applied different cutoffs on the mode
inertia to select a different number of g-dominated modes in the
spectrum. Typically, we have created four sets: one with all the
dipole modes (without any restriction on mode inertia), one with only
three or four most p-dominated modes in each band (corresponding to an
inertia cutoff close to the minima), and two others with intermediate
numbers of dipole modes. Since the observed amplitude of dipole modes
are related to the mode inertia \citep{Dupret09}, selection based on
the inertia would mimic the observed spectra with varying detection
limits based on mode amplitudes. 

This selection is illustrated in Fig.~\ref{fig:model2_inertia_cut} for
Model~2 for a case with 5 radial orders. The results for the four
different choices of the mode set, in decreasing order of inertia
cutoffs are shown by the histograms in the figure. When the full set
of modes is used, there is only one dominant peak in the histogram,
very close to the actual value of \dpg\ (middle left panel of
Fig.~\ref{fig:model2_inertia_cut}). This is because the presence of
the g-dominated modes forming the vertical ridge in the \ped\ have a
large influence in constraining the \dpg. However, for the same
reason, in some realisations, when a few of these crucial frequencies
are perturbed significantly from their true values, they influence the
estimate of \dpg\ strongly and alternative values of \dpg\ are found.
These appear as smaller peaks in the histogram. When an inertia cutoff
is applied  so that the most g-dominated modes are omitted, the
estimation of \dpg\ becomes less stable and a secondary peak at a
lower value is found, although the median value still lies very close
to the actual value.  However, on further lowering the inertia cutoff,
the value of \dpg\ actually stabilises again. This is because the
vertical ridge is now mainly determined by the two dipole modes
closest to the radial mode in each order and the condition for
symmetric distribution of the p-dominated modes becomes more important
to find \dpg.

For each of these artificially created samples of modes containing
radial and dipole modes we have considered four to six different
values for the uncertainty in the frequencies. Throughout this
exercise, we have considered uniform uncertainties in all the mode
frequencies. Thus for each of the three models, we have carried out
the Monte Carlo exercise for nearly 60 different data sets.

The uncertainties in \dpg\ for different levels of uncertainties in
frequencies for the three models and the three observed \kep\ red
giants are given in
Tables~\ref{tab:error_table_model1}--\ref{tab:error_table_obs}.
As an illustration of the exercise, in 
Fig.~\ref{fig:model2_histo} we show the histograms for the
\dpg\ values found by GPS for different datasets constructed
from the theoretical frequencies of Model~2, as described
above. 

\myfigure{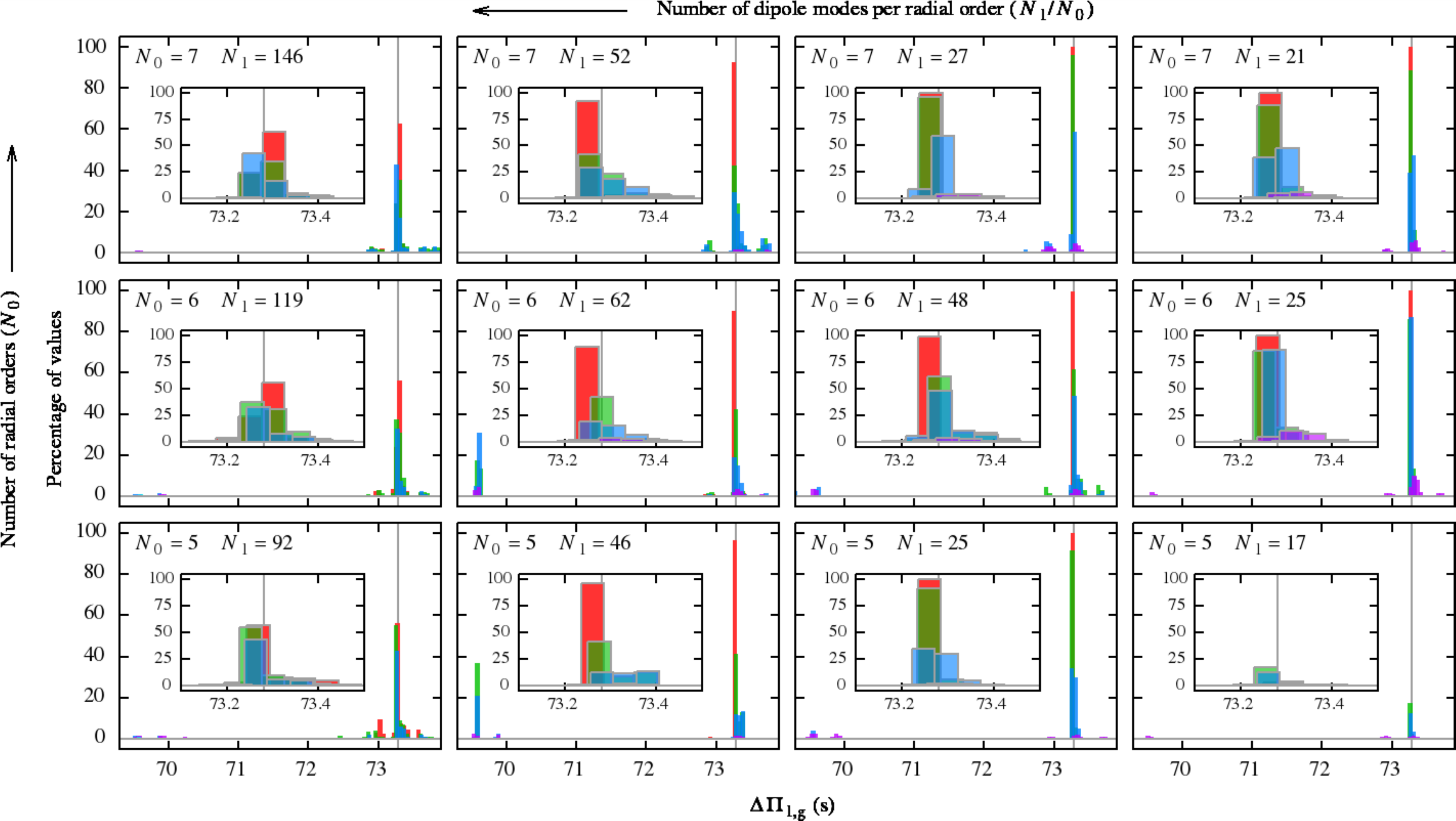}%
{%
Results of the Monte Carlo exercise with different uncertainties
assumed for the frequencies of a $1\msun$ model (Model 2) are shown.
In each panel the histograms of the distribution of \dpg\ determined
by GPS are shown for assumed uncertainties of 0.001\,\muHz\ (in red),
0.005\,\muHz\ (in green), 0.010\,\muHz\ (in blue), 0.050\,\muHz\ (in
magenta) and 0.100\,\muHz\ (in cyan). The different panels show the
different data sets considered for the exercise in terms of number of
radial orders ($N_0$) and number of dipole modes ($N_1$). The grey
vertical line shows the \dpg\ obtained with the full set of
unperturbed frequencies of the model. The insets in each panel show
the main peaks of the histograms. Numerical estimates of \dpg\ from
this exercise are given in Table~\ref{tab:error_table_model2}.
}%
{fig:model2_histo}%
{*}

For each of the mode sets described above for the three models, we
applied GPS after perturbing the frequencies with random uncertainties
corresponding to normal distributions with widths of $0.001,\ 0.005,\
0.010,\ 0.050$ and $0.100$\,\muHz. The results for the three models
show similar trends for uncertainty in \dpg. For the full set of
frequencies the large number of g-dominated modes ensures that \dpg\
is estimated to be very close to the asymptotic value for small
frequency uncertainties. For the smallest frequency uncertainty of
$0.001$\,\muHz, typically almost all the realisations produce a unique
value of \dpg. However, as the uncertainty in frequencies increases,
the probability of obtaining the correct value of \dpg\ decreases,
which is manifested in the appearance of multiple peaks in the
histogram of \dpg\ values. For the cases where we choose fewer number
of modes, namely, only those with lower inertia, the probability of
obtaining a value close to the correct value decreases with decreasing
number of modes, before again increasing for the cases with the least
number of modes. A closer inspection of the situation reveals that the
apparent decrease of the probability of the correct value being
obtained is due to appearance of a few values very close to the
correct one but different by more than one bin in the histogram.  If
one would choose a larger binwidth of the histogram of \dpg, one would
regain the correct value with a larger uncertainty, although not all
values within the range quoted by the $1\sigma$ uncertainties would be
actually permissible. Finally, the increase in probability of finding
the correct value of \dpg\ with fewer modes is essentially
due to the comparatively larger effect of the p-dominated modes in
constraining \dpg\ through the symmetrical distribution criterion, as
compared to the determination of the vertical ridge from the
g-dominated modes. However, if the number of modes per band falls
below three, the determination of \dpg\ becomes unreliable even at
small frequency uncertainties (see
Tables~\ref{tab:error_table_model1}--\ref{tab:error_table_model3}).  

Although the trend of the uncertainties in \dpg\ with increasing
frequency uncertainties is broadly similar in the three models, there
are clear differences in the details. In general, the most stable
determination of \dpg\ happens for Model~1, which is the youngest
among the three models.  For this model, the density of g-modes is
least, and one can reliably estimate \dpg\ even at frequency
uncertainties up to 0.100\,\muHz\ in several cases. For Model~2, \dpg\
is correctly estimated up to frequency uncertainties of 0.050\,\muHz,
while for Model~3 GPS fails to find a reliable value of \dpg\ even at
frequency uncertainties of 0.010\,\muHz\ in some cases. The reason
behind this behaviour can be understood in the following way.  The
coupling between the modes trapped in the envelope and those in the
core decreases as the star evolves up the red-giant branch. This
implies that the number of dipole modes with low inertia decreases
with age \citep{Dupret09}. 

For the method used by GPS, a stronger coupling is beneficial for the
determination of \dpg. In case of stronger coupling the transition
from p-dominated modes in one band to p-dominated modes in the next
band through the g-dominated dipole modes around the location of a
radial mode is less steep, i.e. the g-dominated modes cover a larger
range of inertia. In GPS the determination of \dpg\ depends crucially
on the mixed dipole modes closest to the radial mode at the boundary
of a band through the weighing factors in Eq.~(\ref{eq:pirgi}). In
more evolved models with weaker coupling (e.g., Model~3), small
perturbations in the dipole frequencies closest to the radial modes
have large impact on the determination of  \dpg\ thanks to the steep
slope originating from the weak coupling. On the other hand, when
fewer modes are chosen based on mode inertia the dipole modes closest
to the radial modes are separated by too large a frequency interval,
again providing difficulties for GPS to determine \dpg\ reliably. In
cases with stronger coupling (e.g., Model~1) the less steep transition
of the p-dominated regions reduces the influence of small
perturbations in the dipole frequencies, thus providing a more robust
determination of \dpg. In this case, GPS works even for relatively
higher frequency uncertainties.

In the cases of the three observed \kep\ stars we examine the results
of the Monte Carlo exercise with up to five times the nominal
$1\sigma$ uncertainty in the frequencies ($=0.022\,\muHz$) determined
from the peakbagging exercise. These results are shown in
Table~\ref{tab:error_table_obs}. We find that for \kiconezero\ even
when the uncertainty in the frequencies is increased up to $5\sigma$,
we obtain values consistent with $\dpg = 81.54^{+0.06}_{-0.04}$\,s, as
found with the $1\sigma$ uncertainties, albeit with decreasing
probabilities. This is consistent with the trend shown by Model~1,
which has a very similar \dpg\ value as this star. For the star
\kicnineone\ which is located higher in the red-giant branch, the
results remain consistent up to $2\sigma$ uncertainties in
frequencies. In this case, a secondary peak at $77.72$\,s, which is
only $0.7$\,s seconds away from the highest peak, is found even at
small frequency perturbations. In the case of the further evolved star
\kicfiveeight, GPS is unable to determine \dpg\ at all when the
frequency uncertainties are beyond $3\sigma$. The last two cases are
similar in behaviour to Models~2 and 3, respectively. 

In general, we find that when the sum of the probabilities of the
three highest peaks in the histogram is less than around 75\%, the
value of \dpg\ determined by GPS is not very reliable. In such cases,
we essentially have a number of closely spaced distinct values of
\dpg\ which have comparable probabilities. In many cases it might be
possible to quote a median value of \dpg\ if we use a larger binwidth,
leading to a larger uncertainty value.

\subsection{Iteration between peakbagging and GPS}
\label{sec:appl_peak}

Peakbagging is the craft of finding, identifying and fitting
oscillation modes in a Fourier power spectrum of an oscillating star.
To find real oscillation signal, statistical tests are often applied
\citep[e.g.][and references therein]{Hekker10,Appourchaux14}. These statistical tests essentially provide
a probability of a feature being due to noise or signal. Depending on
the threshold used, this implies that a fraction of the features
selected to be signal can actually be due to noise, and the other way
around; features that are actual signal are not selected. Missing
information is often less harmful than wrong information and hence
making sure that all signal features are indeed due to genuine stellar
oscillations has priority. However, a larger number of observed
oscillation modes could provide additional information and
constraints, required to draw inferences.
\myfiguretwobyone%
{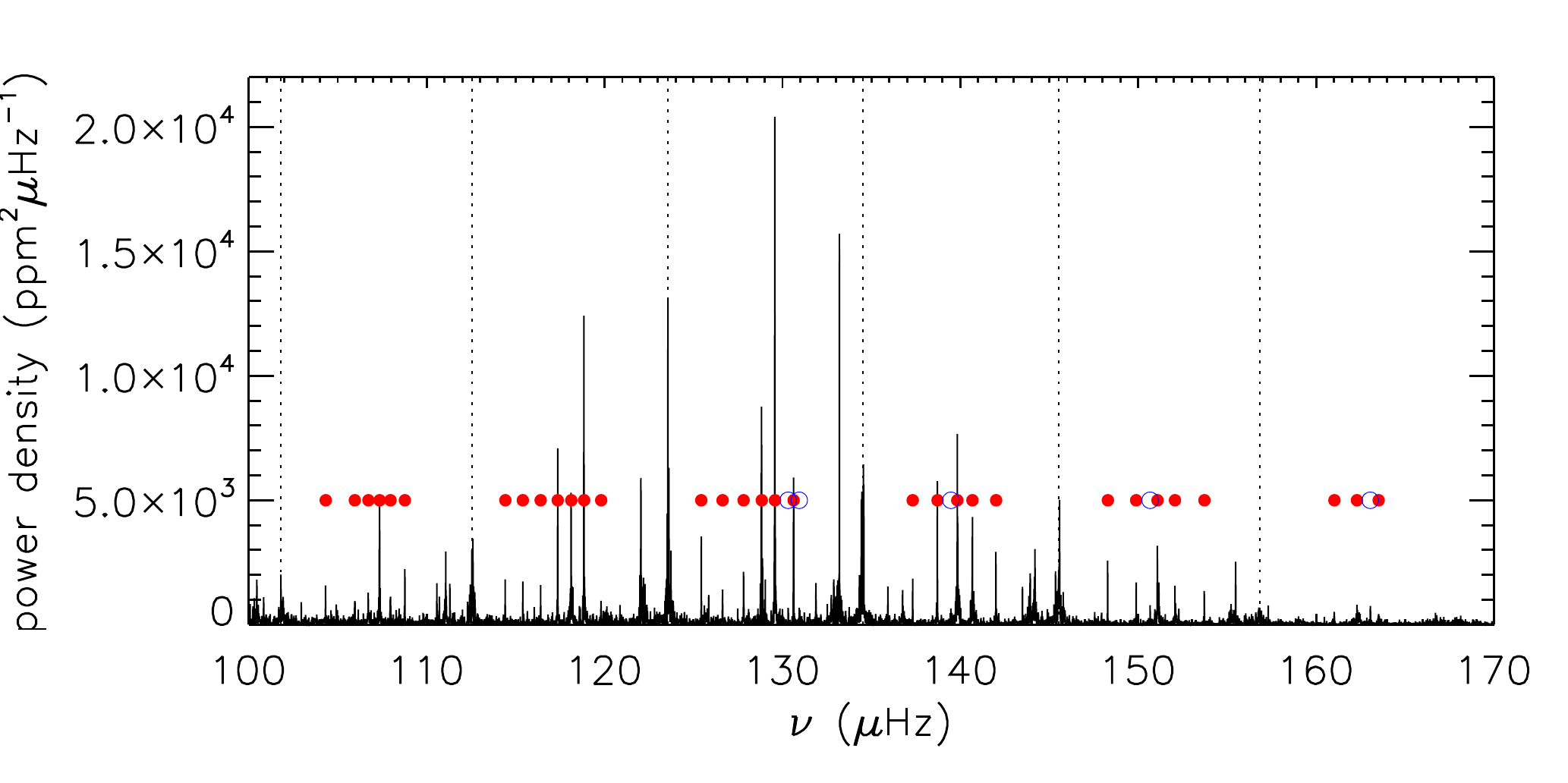}%
{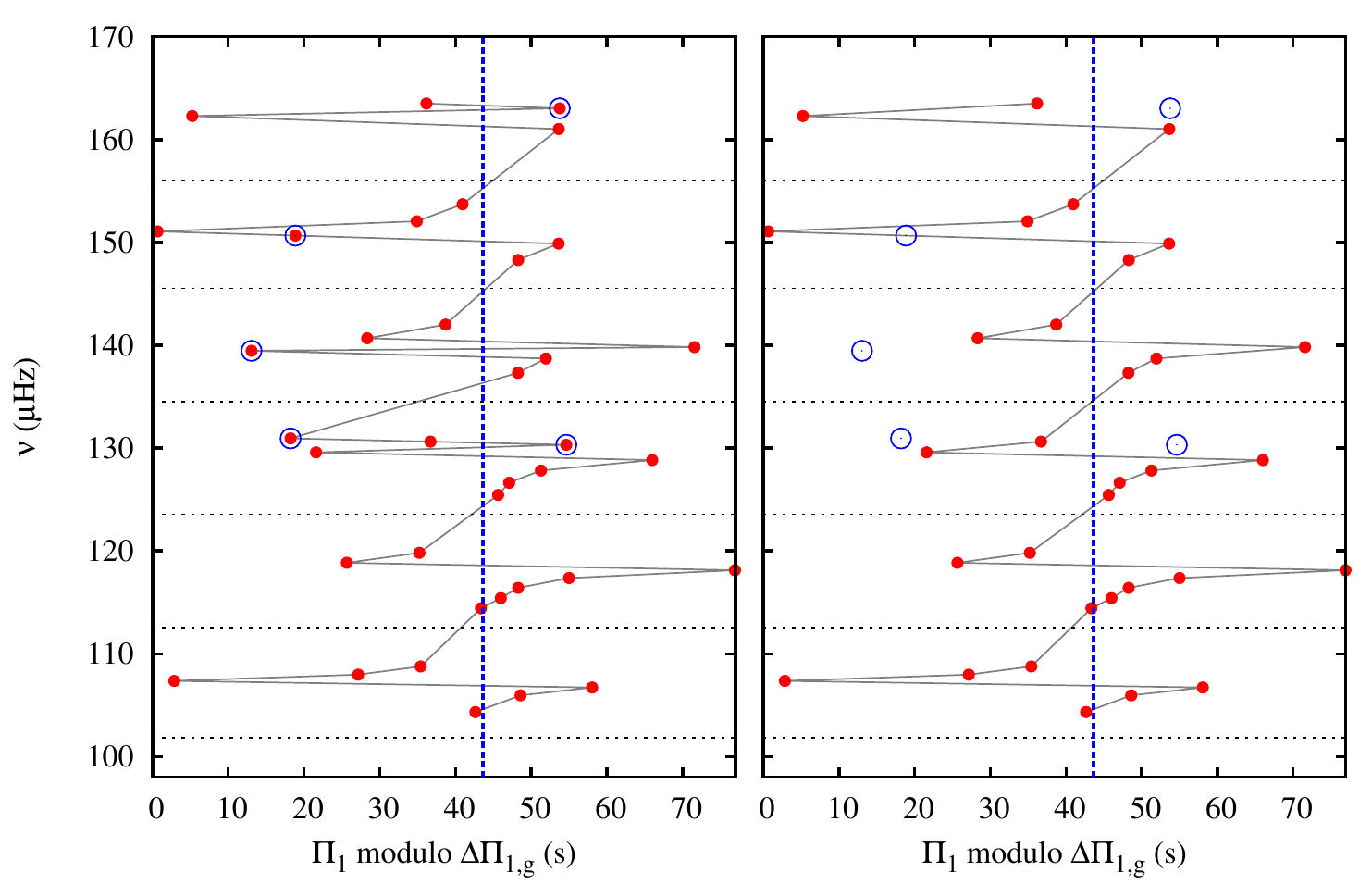}%
{%
The iterative process of using GPS to improve mode detection and
identification is illustrated for the \kep\ red-giant star
\kicnineone.  The top panel shows the power spectrum where the radial
modes are indicated by dotted vertical lines and the dipole modes are
marked by red dots. The dipole modes which were deemed to have been
misidentified based on input from GPS are encircled in blue. The
bottom two panels show the \ped s, where the lines and symbols have
the same meaning as in Fig.~\ref{fig:echelle}. In the bottom left
panel is the diagram using all the initially identified modes, with
the suspected misidentifications encircled in blue. The bottom right
panel shows the \ped\ where the misidentified modes (empty blue
circles) have been excluded from the analysis.
}%
{fig:echelle_ps}%
{}%

When the oscillation signals have been detected, they need to be
identified in terms of radial order, spherical degree and azimuthal
order. Methods such as the universal pattern \citep{Mosser11,Mosser12}
are developed for this purpose.  Additionally, for solar-like
oscillations accurate frequencies, mode widths and heights with
uncertainties can be obtained from Lorentzian fits to the oscillation
signals in the Fourier power spectrum. This information is of
importance when using the oscillation modes to compute intrinsic
stellar properties and infer the internal stellar structure of a star.
For GPS the frequencies and mode degree are the most important
observational inputs to compute the period spacing. We note here that
GPS can, in principle, work for any azimuthal order ($m$) as long as
the azimuthal order is the same for all modes. Here it is, however,
always applied to modes with $m=0$.

In case GPS cannot constrain the period spacing it is possible to
identify  features in the Fourier power spectrum that do not conform
to the expectations. For these features visual checks are performed in
the Fourier power spectrum to gauge whether the feature is a genuine
oscillation mode, its identification is correct (in this way we can for instance identify a mode with a different azimuthal order) and/or if the
frequency is accurately determined. In this way it is possible to
iterate between peakbagging and GPS to make sure that correctly
identified true signals are used for the determination of the period
spacing.

An example of this iteration process is shown in
Fig.~\ref{fig:echelle_ps} where both the Fourier power spectrum of
\kicnineone\ and the corresponding \ped s are shown. The blue and red
dots in the Fourier power spectrum (top panel of
Fig.~\ref{fig:echelle_ps}) indicate the initially detected $l=1$
modes. These frequencies did not provide a proper \ped\ (bottom left
panel of Fig.~\ref{fig:echelle_ps}). Using the information of the
problematic frequencies and the computed approximate period spacing we
have been able to optimise the detection and identification of the
$l=1$ modes. In this way we discarded incorrectly identified features
of the power spectrum and identified more modes over a wider frequency
range (red dots in top panel and right bottom panel of
Fig.~\ref{fig:echelle_ps}), which allowed GPS to determine the period
spacing of \kicnineone\ to be $76.98 \pm 0.03$\,s.

\section{Conclusion}
\label{sec:concl}

We have devised a new method (GPS, g-mode period spacing finder) to
estimate the period spacing of g-modes in red-giant stars. Such modes
can usually not be detected in the observed Fourier spectrum of a red
giant because of their low amplitudes. However, the period spacing of
these modes is related to the conditions in the core, specifically,
the buoyancy frequency. Therefore, if one is able to determine the
g-mode period spacing from the observed modes, it may be used as a
strong constraint in finding a suitable model for a star.

The automated method devised here is based on the \ped\ and
essentially seeks the period spacing for which a vertical alignment of
the gravity-dominated modes is present as well as a symmetric
distribution of the pressure-dominated modes. The method has been
extensively tested on model frequencies. For each model different sets
of frequencies have been selected based on their inertia, different
frequency ranges have been taken and different (uniform) uncertainties
on the frequencies have been tested in a Monte Carlo approach.  This
shows that often more than one period spacing provides a possible
answer. These period spacing values are distinct with ranges without
possible solutions separating them. Therefore, we provide the three
most probable period spacings for each investigated case (both models
and observations) with their probability and uncertainty.  For both
the models and observations we obtain period spacings with high
probability when the uncertainties on the frequencies are in the range
0.01--0.05\,$\mu$Hz or below. In case of the models these are indeed
consistent with the asymptotic period spacings.

Although, GPS is only applied to red-giant branch models and stars in
this work, it has already been applied successfully to a red-clump
star by \citet{Silva14}. The potential of GPS for red-clump stars will
be explored more extensively elsewhere.

Finally, GPS has significant potential once the extraction of mixed
modes of large numbers of stars is possible. First attempts for this
are already present such as the work by \citet{Stello13} and \citet{Mosser12}.  However, determinations of accurate individual frequencies for a large number of stars are currently not available and hence such analysis is beyond the scope of this paper. Work to develop tools to get these accurate individual frequencies is underway and application of GPS to such a large sample will be published in forthcoming publications.
\section{Acknowledgments}

AD, AM and UG acknowledge support from the National Initiative on
Undergraduate Science (NIUS) programme of HBCSE (TIFR).  SH
acknowledges financial support from the Netherlands Organisation for
Scientific Research (NWO).  The research leading to the presented
results has received funding from the European Research Council under
the European Community's Seventh Framework Programme (FP7/2007-2013) /
ERC grant agreement no 338251 (StellarAges).

\appendix

\section{Detailed results of Monte Carlo exercise}

\begin{table*}
\caption{%
Results of a Monte Carlo exercise with a $1\msun$ red giant model at
an age of 12.4\,Gyr (Model~1).  $N_0$ and $N_1$ are the number of
$l=0$ and $l=1$ modes, respectively, that are used in the exercise.
$\delta\nu$ is the uncertainty introduced in the frequencies and $p$
is the probability of the associated \dpg\ value.  The asymptotic
value of \dpg\ for this model is $82.61$\,s.
\label{tab:error_table_model1}
}
\renewcommand{\arraystretch}{1.5}
{\footnotesize
\begin{center}
\begin{tabular}{cccccccccccc}
\hline
$N_0$  & $N_1$ & \multicolumn{10}{c}{$\delta\nu$\,($\mu$Hz)} \\
\cline{3-12}
(Range &        & \multicolumn{2}{c}{0.001}         & \multicolumn{2}{c}{0.005}         & \multicolumn{2}{c}{0.010}         & \multicolumn{2}{c}{0.050}         & \multicolumn{2}{c}{0.100}        \\
\cline{3-12}
in $\mu$Hz) &       & $p$ & \dpg\,(s)  & $p$ & \dpg\,(s)  & $p$ & \dpg\,(s)  & $p$ & \dpg\,(s)  & $p$ & \dpg\,(s)   \\     
\hline
   6   &   60  & 1.0000 & $82.48_{-0.01}^{+0.01}$  & 0.8263 & $82.49_{-0.02}^{+0.02}$  & 0.5762 & $82.50_{-0.03}^{+0.15}$  & 0.1546 & $84.82_{-0.06}^{+0.08}$  & 0.1778 & $85.97_{-0.26}^{+0.33}$ \\
(75.00)&       & \blank &      \blank              & 0.1008 & $82.62_{-0.08}^{+0.13}$  & 0.2785 & $83.34_{-0.03}^{+0.07}$  & 0.1379 & $83.45_{-0.07}^{+0.07}$  & 0.1127 & $87.11_{-0.08}^{+0.08}$ \\
       &       & \blank &      \blank              & 0.0404 & $83.34_{-0.02}^{+0.02}$  & 0.0932 & $81.70_{-0.20}^{+0.05}$  & 0.1208 & $87.11_{-0.03}^{+0.04}$  & 0.0794 & $83.53_{-0.11}^{+0.11}$ \\
\cline{3-12}
       &   32  & 1.0000 & $82.48_{-0.01}^{+0.01}$  & 1.0000 & $82.47_{-0.01}^{+0.01}$  & 0.9974 & $82.47_{-0.02}^{+0.03}$  & 0.6605 & $82.50_{-0.04}^{+0.03}$  & 0.4337 & $82.50_{-0.07}^{+0.09}$ \\
       &       & \blank &      \blank              & \blank &      \blank              & 0.0019 & $82.62_{-0.04}^{+0.05}$  & 0.1717 & $83.33_{-0.07}^{+0.10}$  & 0.2104 & $83.42_{-0.13}^{+0.10}$ \\
       &       & \blank &      \blank              & \blank &      \blank              & 0.0003 & $81.51_{-0.02}^{+0.02}$  & 0.0736 & $81.56_{-0.09}^{+0.17}$  & 0.1167 & $84.64_{-0.17}^{+0.21}$ \\
\cline{3-12}
       &   20  & 1.0000 & $81.71_{-0.01}^{+0.01}$  & 0.9899 & $81.70_{-0.02}^{+0.02}$  & 0.8760 & $81.73_{-0.04}^{+0.03}$  & 0.4986 & $82.51_{-0.04}^{+0.03}$  & 0.4880 & $82.51_{-0.04}^{+0.05}$ \\
       &       & \blank &      \blank              & 0.0095 & $82.50_{-0.03}^{+0.02}$  & 0.1240 & $82.50_{-0.02}^{+0.02}$  & 0.4345 & $81.73_{-0.03}^{+0.04}$  & 0.2798 & $81.72_{-0.11}^{+0.08}$ \\
       &       & \blank &      \blank              & 0.0003 & $82.55_{-0.02}^{+0.02}$  & \blank &      \blank              & 0.0129 & $68.98_{-0.02}^{+0.02}$  & 0.0353 & $68.96_{-0.04}^{+0.04}$ \\
\cline{3-12}
       &   17  & 1.0000 & $81.71_{-0.01}^{+0.01}$  & 0.9999 & $81.71_{-0.02}^{+0.02}$  & 0.9899 & $81.70_{-0.02}^{+0.03}$  & 0.8062 & $81.74_{-0.03}^{+0.05}$  & 0.5690 & $81.76_{-0.06}^{+0.06}$ \\
       &       & \blank &      \blank              & 0.0001 & $81.76_{-0.02}^{+0.02}$  & 0.0101 & $81.80_{-0.02}^{+0.02}$  & 0.1782 & $82.54_{-0.03}^{+0.04}$  & 0.3188 & $82.54_{-0.05}^{+0.05}$ \\
       &       & \blank &      \blank              & \blank &      \blank              & \blank &      \blank              & 0.0092 & $81.88_{-0.02}^{+0.02}$  & 0.0362 & $71.24_{-0.08}^{+0.07}$ \\
\hline
   6   &   48  & 1.0000 & $82.48_{-0.01}^{+0.01}$  & 0.8573 & $82.49_{-0.04}^{+0.03}$  & 0.5375 & $82.50_{-0.04}^{+0.15}$  & 0.3078 & $87.24_{-0.03}^{+0.03}$  & 0.1401 & $87.19_{-0.15}^{+0.21}$ \\
(62.50)&       & \blank &      \blank              & 0.0820 & $82.63_{-0.07}^{+0.08}$  & 0.2217 & $83.16_{-0.02}^{+0.02}$  & 0.1224 & $83.59_{-0.06}^{+0.05}$  & 0.1352 & $85.73_{-0.12}^{+0.22}$ \\
       &       & \blank &      \blank              & 0.0479 & $83.16_{-0.02}^{+0.02}$  & 0.1741 & $83.34_{-0.03}^{+0.16}$  & 0.1214 & $88.15_{-0.03}^{+0.05}$  & 0.0854 & $88.15_{-0.06}^{+0.12}$ \\
\cline{3-12}
       &   24  & 1.0000 & $81.71_{-0.01}^{+0.01}$  & 0.9650 & $81.74_{-0.02}^{+0.02}$  & 0.8157 & $81.73_{-0.02}^{+0.02}$  & 0.3833 & $82.50_{-0.04}^{+0.04}$  & 0.2573 & $82.53_{-0.07}^{+0.12}$ \\
       &       & \blank &      \blank              & 0.0300 & $81.69_{-0.02}^{+0.02}$  & 0.1005 & $72.74_{-0.02}^{+0.03}$  & 0.2788 & $81.73_{-0.05}^{+0.14}$  & 0.1292 & $83.46_{-0.10}^{+0.10}$ \\
       &       & \blank &      \blank              & 0.0039 & $72.74_{-0.02}^{+0.02}$  & 0.0265 & $81.68_{-0.02}^{+0.02}$  & 0.0946 & $83.45_{-0.05}^{+0.09}$  & 0.1211 & $72.70_{-0.08}^{+0.07}$ \\
\cline{3-12}
       &   18  & 0.9987 & $82.47_{-0.02}^{+0.02}$  & 0.6545 & $82.46_{-0.02}^{+0.02}$  & 0.5449 & $82.48_{-0.02}^{+0.02}$  & 0.4711 & $82.48_{-0.03}^{+0.04}$  & 0.3255 & $82.48_{-0.08}^{+0.07}$ \\
       &       & 0.0013 & $81.72_{-0.02}^{+0.02}$  & 0.2696 & $81.71_{-0.02}^{+0.02}$  & 0.3281 & $81.73_{-0.02}^{+0.02}$  & 0.1890 & $72.71_{-0.06}^{+0.06}$  & 0.1425 & $72.69_{-0.07}^{+0.07}$ \\
       &       & \blank &      \blank              & 0.0557 & $72.72_{-0.02}^{+0.02}$  & 0.1204 & $72.73_{-0.02}^{+0.02}$  & 0.1821 & $81.72_{-0.03}^{+0.04}$  & 0.0898 & $71.91_{-0.06}^{+0.08}$ \\
\cline{3-12}
       &   13  & 1.0000 & $81.71_{-0.01}^{+0.01}$  & 0.9994 & $81.73_{-0.04}^{+0.02}$  & 0.9468 & $81.72_{-0.02}^{+0.02}$  & 0.5291 & $81.70_{-0.07}^{+0.05}$  & 0.3462 & $81.70_{-0.12}^{+0.09}$ \\
       &       & \blank &      \blank              & 0.0006 & $71.24_{-0.02}^{+0.02}$  & 0.0379 & $71.23_{-0.02}^{+0.04}$  & 0.1957 & $71.26_{-0.02}^{+0.02}$  & 0.1449 & $71.27_{-0.04}^{+0.04}$ \\
       &       & \blank &      \blank              & \blank &      \blank              & 0.0151 & $81.67_{-0.02}^{+0.02}$  & 0.1352 & $61.97_{-0.02}^{+0.04}$  & 0.1117 & $61.97_{-0.03}^{+0.04}$ \\
\hline
   4   &   39  & 0.9995 & $81.71_{-0.02}^{+0.02}$  & 0.5776 & $82.49_{-0.04}^{+0.03}$  & 0.6377 & $82.49_{-0.02}^{+0.05}$  & 0.1418 & $83.49_{-0.15}^{+0.21}$  & 0.0950 & $86.86_{-0.22}^{+0.16}$ \\
(50.23)&       & 0.0005 & $82.46_{-0.02}^{+0.02}$  & 0.4008 & $81.71_{-0.02}^{+0.02}$  & 0.1251 & $83.19_{-0.02}^{+0.02}$  & 0.1314 & $85.84_{-0.06}^{+0.25}$  & 0.0700 & $79.24_{-0.26}^{+0.41}$ \\
       &       & \blank &      \blank              & 0.0176 & $82.62_{-0.07}^{+0.12}$  & 0.1011 & $81.73_{-0.03}^{+0.02}$  & 0.1035 & $84.37_{-0.07}^{+0.08}$  & 0.0331 & $77.49_{-0.05}^{+0.05}$ \\
\cline{3-12}
       &   26  & 1.0000 & $81.71_{-0.01}^{+0.01}$  & 0.9997 & $81.72_{-0.02}^{+0.02}$  & 0.9296 & $81.73_{-0.04}^{+0.02}$  & 0.4497 & $82.50_{-0.04}^{+0.11}$  & 0.2048 & $82.50_{-0.19}^{+0.15}$ \\
       &       & \blank &      \blank              & 0.0002 & $73.32_{-0.02}^{+0.02}$  & 0.0436 & $82.49_{-0.02}^{+0.02}$  & 0.1565 & $83.45_{-0.05}^{+0.07}$  & 0.1943 & $83.43_{-0.13}^{+0.11}$ \\
       &       & \blank &      \blank              & 0.0001 & $82.47_{-0.02}^{+0.02}$  & 0.0149 & $73.33_{-0.03}^{+0.02}$  & 0.0941 & $84.35_{-0.05}^{+0.05}$  & 0.1210 & $84.37_{-0.07}^{+0.08}$ \\
\cline{3-12}
       &   22  & 1.0000 & $81.71_{-0.01}^{+0.01}$  & 1.0000 & $81.71_{-0.01}^{+0.01}$  & 1.0000 & $81.71_{-0.03}^{+0.04}$  & 0.4267 & $81.72_{-0.04}^{+0.03}$  & 0.3023 & $82.52_{-0.06}^{+0.08}$ \\
       &       & \blank &      \blank              & \blank &      \blank              & \blank &      \blank              & 0.3280 & $82.49_{-0.03}^{+0.04}$  & 0.1248 & $83.45_{-0.11}^{+0.10}$ \\
       &       & \blank &      \blank              & \blank &      \blank              & \blank &      \blank              & 0.0321 & $56.75_{-0.03}^{+0.08}$  & 0.1160 & $81.72_{-0.15}^{+0.11}$ \\
\cline{3-12}
       &   14  & 1.0000 & $81.71_{-0.01}^{+0.01}$  & 0.9999 & $81.73_{-0.04}^{+0.03}$  & 0.9529 & $81.72_{-0.04}^{+0.03}$  & 0.2830 & $81.73_{-0.03}^{+0.04}$  & 0.1321 & $82.49_{-0.04}^{+0.05}$ \\
       &       & \blank &      \blank              & 0.0001 & $71.24_{-0.02}^{+0.02}$  & 0.0253 & $71.24_{-0.02}^{+0.02}$  & 0.1409 & $66.12_{-0.03}^{+0.02}$  & 0.1072 & $66.10_{-0.05}^{+0.04}$ \\
       &       & \blank &      \blank              & \blank &      \blank              & 0.0156 & $81.89_{-0.02}^{+0.02}$  & 0.1089 & $61.98_{-0.02}^{+0.02}$  & 0.0861 & $61.97_{-0.03}^{+0.04}$ \\
\hline
\end{tabular}
\end{center}
}
\end{table*}

\begin{table*}
\caption{%
Results of a Monte Carlo exercise with a $1\msun$ red giant model at
an age of 12.5\,Gyr (Model~2).  The quantities shown have the same
meaning as in Table~\ref{tab:error_table_model1}.  The asymptotic
value of \dpg\ for this model is $73.49$\,s.
\label{tab:error_table_model2}
}
\renewcommand{\arraystretch}{1.5}
{\footnotesize
\begin{center}
\begin{tabular}{cccccccccccc}
\hline
$N_0$  & $N_1$ & \multicolumn{10}{c}{$\delta\nu$\,($\mu$Hz)} \\
\cline{3-12}
(Range &       & \multicolumn{2}{c}{0.001}         & \multicolumn{2}{c}{0.005}         & \multicolumn{2}{c}{0.010}         & \multicolumn{2}{c}{0.050}         & \multicolumn{2}{c}{0.100}        \\
\cline{3-12}
in $\mu$Hz) &       & $p$ & \dpg\,(s)  & $p$ & \dpg\,(s)  & $p$ & \dpg\,(s)  & $p$ & \dpg\,(s)  & $p$ & \dpg\,(s)     \\     
\hline

   7   &  146  & 0.9074 & $73.30_{-0.04}^{+0.02}$  & 0.6699 & $73.29_{-0.04}^{+0.03}$  & 0.6449 & $73.27_{-0.03}^{+0.05}$  & 0.0688 & $78.72_{-0.06}^{+0.06}$  & 0.2578 & $66.23_{-1.83}^{+1.44}$ \\
(46.19)&       & 0.0256 & $72.90_{-0.02}^{+0.02}$  & 0.0937 & $72.92_{-0.05}^{+0.10}$  & 0.1122 & $73.76_{-0.13}^{+0.09}$  & 0.0603 & $78.20_{-0.06}^{+0.06}$  & 0.0307 & $68.43_{-0.07}^{+0.07}$ \\
       &       & 0.0218 & $73.40_{-0.02}^{+0.02}$  & 0.0417 & $74.16_{-0.03}^{+0.08}$  & 0.0338 & $74.37_{-0.02}^{+0.08}$  & 0.0371 & $68.84_{-0.04}^{+0.03}$  & 0.0302 & $68.84_{-0.07}^{+0.06}$ \\
\cline{3-12}
       &   52  & 0.9263 & $73.25_{-0.02}^{+0.02}$  & 0.7389 & $73.27_{-0.03}^{+0.05}$  & 0.6214 & $73.29_{-0.04}^{+0.06}$  & 0.1104 & $85.35_{-0.05}^{+0.04}$  & 0.0319 & $85.37_{-0.09}^{+0.07}$ \\
       &       & 0.0583 & $73.33_{-0.04}^{+0.07}$  & 0.1364 & $73.69_{-0.06}^{+0.03}$  & 0.1657 & $73.69_{-0.04}^{+0.07}$  & 0.0541 & $68.81_{-0.05}^{+0.05}$  & 0.0251 & $84.78_{-0.07}^{+0.08}$ \\
       &       & 0.0130 & $72.94_{-0.05}^{+0.02}$  & 0.0967 & $72.90_{-0.04}^{+0.03}$  & 0.0672 & $72.86_{-0.03}^{+0.04}$  & 0.0527 & $69.19_{-0.05}^{+0.04}$  & 0.0218 & $85.95_{-0.06}^{+0.08}$ \\
\cline{3-12}
       &   27  & 1.0000 & $73.26_{-0.01}^{+0.01}$  & 0.9590 & $73.26_{-0.02}^{+0.02}$  & 0.6772 & $73.29_{-0.02}^{+0.02}$  & 0.1087 & $73.32_{-0.04}^{+0.06}$  & 0.5955 & $55.33_{-3.94}^{+5.43}$ \\
       &       & \blank & $     \blank          $  & 0.0176 & $72.91_{-0.02}^{+0.02}$  & 0.1148 & $72.92_{-0.04}^{+0.04}$  & 0.1014 & $72.93_{-0.06}^{+0.06}$  & 0.0222 & $68.89_{-0.10}^{+0.33}$ \\
       &       & \blank & $     \blank          $  & 0.0100 & $73.34_{-0.04}^{+0.06}$  & 0.0512 & $77.04_{-0.02}^{+0.03}$  & 0.0719 & $77.55_{-0.06}^{+0.06}$  & 0.0184 & $73.34_{-0.06}^{+0.06}$ \\
\cline{3-12}
       &   21  & 1.0000 & $73.26_{-0.01}^{+0.01}$  & 0.9973 & $73.26_{-0.02}^{+0.02}$  & 0.8618 & $73.28_{-0.04}^{+0.03}$  & 0.1412 & $73.32_{-0.05}^{+0.04}$  & 0.0237 & $85.06_{-0.08}^{+0.09}$ \\
       &       & \blank & $     \blank          $  & 0.0023 & $76.00_{-0.03}^{+0.02}$  & 0.0649 & $76.00_{-0.02}^{+0.02}$  & 0.0562 & $85.11_{-0.06}^{+0.06}$  & 0.0214 & $78.51_{-0.07}^{+0.07}$ \\
       &       & \blank & $     \blank          $  & 0.0003 & $65.76_{-0.02}^{+0.02}$  & 0.0215 & $73.35_{-0.02}^{+0.02}$  & 0.0445 & $72.93_{-0.05}^{+0.05}$  & 0.0187 & $73.33_{-0.06}^{+0.06}$ \\
\hline
   5   &  119  & 0.8834 & $73.29_{-0.04}^{+0.03}$  & 0.7940 & $73.28_{-0.04}^{+0.04}$  & 0.4460 & $73.28_{-0.02}^{+0.05}$  & 0.0502 & $77.77_{-0.07}^{+0.07}$  & 0.0271 & $69.20_{-0.06}^{+0.05}$ \\
(38.65)&       & 0.0686 & $72.98_{-0.05}^{+0.04}$  & 0.0681 & $73.00_{-0.13}^{+0.04}$  & 0.0341 & $59.97_{-0.02}^{+0.02}$  & 0.0486 & $68.83_{-0.04}^{+0.03}$  & 0.0226 & $68.84_{-0.05}^{+0.05}$ \\
       &       & 0.0303 & $73.40_{-0.02}^{+0.02}$  & 0.0589 & $73.66_{-0.05}^{+0.06}$  & 0.0332 & $62.72_{-0.02}^{+0.02}$  & 0.0340 & $68.05_{-0.03}^{+0.04}$  & 0.0195 & $69.58_{-0.05}^{+0.05}$ \\
\cline{3-12}
       &   62  & 0.8994 & $73.25_{-0.02}^{+0.02}$  & 0.4248 & $73.28_{-0.02}^{+0.02}$  & 0.4136 & $73.29_{-0.04}^{+0.05}$  & 0.0973 & $68.81_{-0.03}^{+0.04}$  & 0.2813 & $73.87_{-3.03}^{+2.57}$ \\
       &       & 0.0721 & $73.33_{-0.04}^{+0.07}$  & 0.3100 & $69.60_{-0.03}^{+0.04}$  & 0.3547 & $69.60_{-0.02}^{+0.02}$  & 0.0779 & $69.21_{-0.05}^{+0.04}$  & 0.1250 & $79.83_{-1.40}^{+1.37}$ \\
       &       & 0.0251 & $72.89_{-0.05}^{+0.05}$  & 0.0830 & $69.23_{-0.02}^{+0.02}$  & 0.0898 & $69.25_{-0.04}^{+0.02}$  & 0.0765 & $69.58_{-0.04}^{+0.03}$  & 0.0242 & $77.79_{-0.08}^{+0.08}$ \\
\cline{3-12}
       &   48  & 0.9968 & $73.26_{-0.02}^{+0.02}$  & 0.7863 & $73.29_{-0.02}^{+0.04}$  & 0.7139 & $73.29_{-0.03}^{+0.05}$  & 0.0909 & $73.32_{-0.05}^{+0.05}$  & 0.0305 & $84.79_{-0.08}^{+0.08}$ \\
       &       & 0.0032 & $73.36_{-0.05}^{+0.05}$  & 0.0839 & $73.64_{-0.03}^{+0.04}$  & 0.0846 & $69.61_{-0.04}^{+0.03}$  & 0.0709 & $69.57_{-0.03}^{+0.04}$  & 0.0296 & $85.37_{-0.08}^{+0.09}$ \\
       &       & \blank & $     \blank          $  & 0.0613 & $72.89_{-0.02}^{+0.03}$  & 0.0784 & $69.24_{-0.02}^{+0.04}$  & 0.0682 & $68.80_{-0.03}^{+0.04}$  & 0.0256 & $83.05_{-0.07}^{+0.07}$ \\
\cline{3-12}
       &   25  & 1.0000 & $73.26_{-0.01}^{+0.01}$  & 0.9972 & $73.26_{-0.02}^{+0.02}$  & 0.9640 & $73.28_{-0.02}^{+0.02}$  & 0.2482 & $73.32_{-0.05}^{+0.05}$  & 0.0268 & $69.28_{-0.11}^{+0.31}$ \\
       &       & \blank & $     \blank          $  & 0.0027 & $76.71_{-0.02}^{+0.03}$  & 0.0224 & $76.71_{-0.04}^{+0.03}$  & 0.0506 & $68.81_{-0.04}^{+0.04}$  & 0.0180 & $73.34_{-0.07}^{+0.07}$ \\
       &       & \blank & $     \blank          $  & 0.0001 & $73.35_{-0.02}^{+0.02}$  & 0.0034 & $73.22_{-0.02}^{+0.02}$  & 0.0342 & $76.71_{-0.04}^{+0.03}$  & 0.0131 & $68.80_{-0.06}^{+0.08}$ \\
\hline
   5   &   92  & 0.7762 & $73.28_{-0.02}^{+0.06}$  & 0.7104 & $73.26_{-0.02}^{+0.04}$  & 0.5375 & $73.27_{-0.02}^{+0.04}$  & 0.0445 & $77.76_{-0.06}^{+0.07}$  & 0.0494 & $68.67_{-0.26}^{+0.19}$ \\
(30.76)&       & 0.1465 & $73.02_{-0.03}^{+0.03}$  & 0.1069 & $72.93_{-0.10}^{+0.06}$  & 0.0546 & $62.71_{-0.02}^{+0.02}$  & 0.0400 & $68.83_{-0.04}^{+0.04}$  & 0.0266 & $69.20_{-0.07}^{+0.06}$ \\
       &       & 0.0635 & $73.58_{-0.02}^{+0.06}$  & 0.0549 & $73.64_{-0.04}^{+0.09}$  & 0.0437 & $54.89_{-0.04}^{+0.03}$  & 0.0326 & $69.91_{-0.04}^{+0.03}$  & 0.0233 & $69.56_{-0.06}^{+0.06}$ \\
\cline{3-12}
       &   46  & 0.9668 & $73.26_{-0.02}^{+0.02}$  & 0.6239 & $73.29_{-0.03}^{+0.07}$  & 0.3787 & $73.33_{-0.06}^{+0.05}$  & 0.0488 & $69.54_{-0.03}^{+0.04}$  & 0.8130 & $57.41_{-5.12}^{+6.63}$ \\
       &       & 0.0174 & $73.33_{-0.03}^{+0.04}$  & 0.3674 & $69.58_{-0.02}^{+0.02}$  & 0.2182 & $69.58_{-0.02}^{+0.02}$  & 0.0408 & $73.32_{-0.05}^{+0.05}$  & 0.0323 & $70.35_{-0.45}^{+0.71}$ \\
       &       & 0.0145 & $72.91_{-0.02}^{+0.02}$  & 0.0028 & $63.68_{-0.02}^{+0.02}$  & 0.0485 & $54.73_{-0.02}^{+0.02}$  & 0.0320 & $69.88_{-0.03}^{+0.03}$  & 0.0309 & $69.18_{-0.35}^{+0.34}$ \\
\cline{3-12}
       &   25  & 1.0000 & $73.26_{-0.01}^{+0.01}$  & 0.9731 & $73.26_{-0.02}^{+0.02}$  & 0.6901 & $73.27_{-0.03}^{+0.04}$  & 0.0689 & $69.54_{-0.05}^{+0.05}$  & 0.4665 & $58.07_{-5.43}^{+5.20}$ \\
       &       & \blank & $     \blank          $  & 0.0078 & $69.55_{-0.04}^{+0.02}$  & 0.0970 & $67.35_{-0.02}^{+0.02}$  & 0.0525 & $73.32_{-0.06}^{+0.05}$  & 0.3484 & $71.98_{-4.34}^{+5.52}$ \\
       &       & \blank & $     \blank          $  & 0.0034 & $72.86_{-0.02}^{+0.02}$  & 0.0463 & $65.50_{-0.02}^{+0.02}$  & 0.0520 & $69.89_{-0.04}^{+0.05}$  & 0.0090 & $82.59_{-0.09}^{+0.11}$ \\
\cline{3-12}
       &   17  & 0.9565 & $77.13_{-0.02}^{+0.02}$  & 0.4966 & $77.14_{-0.04}^{+0.03}$  & 0.3841 & $77.15_{-0.05}^{+0.05}$  & 0.0563 & $77.13_{-0.06}^{+0.06}$  & 0.0202 & $93.53_{-0.10}^{+0.08}$ \\
       &       & 0.0230 & $73.28_{-0.02}^{+0.02}$  & 0.2832 & $64.14_{-0.04}^{+0.02}$  & 0.3224 & $64.13_{-0.03}^{+0.04}$  & 0.0506 & $77.53_{-0.06}^{+0.08}$  & 0.0180 & $94.19_{-0.10}^{+0.10}$ \\
       &       & 0.0129 & $64.12_{-0.02}^{+0.02}$  & 0.1760 & $73.26_{-0.02}^{+0.02}$  & 0.1585 & $73.27_{-0.02}^{+0.03}$  & 0.0485 & $73.32_{-0.06}^{+0.07}$  & 0.0164 & $99.77_{-0.09}^{+0.11}$ \\
\hline
\end{tabular}
\end{center}
}
\end{table*}

\begin{table*}
\caption{%
Results of a Monte Carlo exercise with a $1\msun$ red giant model at
an age of 12.6\,Gyr (Model~3).  The quantities shown have the same
meaning as in Table~\ref{tab:error_table_model1}.  The asymptotic
value of \dpg\ for this model is $62.15$\,s.
\label{tab:error_table_model3}
}
\renewcommand{\arraystretch}{1.5}
{\footnotesize
\begin{center}
\begin{tabular}{cccccccccc}
\hline
$N_0$  & $N_1$ & \multicolumn{8}{c}{$\delta\nu$\,($\mu$Hz)} \\
\cline{3-10}
(Range &       & \multicolumn{2}{c}{0.001}         & \multicolumn{2}{c}{0.005}         & \multicolumn{2}{c}{0.010}         & \multicolumn{2}{c}{0.050}        \\
\cline{3-10}
in $\mu$Hz) &       & $p$ & \dpg\,(s)  & $p$ & \dpg\,(s)  & $p$ & \dpg\,(s)  & $p$ & \dpg\,(s)  \\     
\hline
   6   &  422  & 0.8710 & $62.05_{-0.02}^{+0.02}$  & 0.0676 & $62.07_{-0.02}^{+0.02}$  & 0.0393 & $53.55_{-0.02}^{+0.02}$  & 0.8152 & $55.48_{-7.24}^{+10.7}$ \\
(27.36)&       & 0.0795 & $62.14_{-0.04}^{+0.03}$  & 0.0272 & $55.87_{-0.02}^{+0.02}$  & 0.0270 & $52.51_{-0.02}^{+0.05}$  & 0.0914 & $78.28_{-2.74}^{+3.51}$ \\
       &       & 0.0309 & $62.37_{-0.03}^{+0.08}$  & 0.0237 & $52.52_{-0.02}^{+0.02}$  & 0.0129 & $50.95_{-0.02}^{+0.02}$  & 0.0062 & $86.74_{-0.25}^{+0.27}$ \\
\cline{3-10}
       &  146  & 0.8552 & $62.06_{-0.02}^{+0.02}$  & 0.2898 & $62.06_{-0.04}^{+0.03}$  & 0.0956 & $62.07_{-0.04}^{+0.12}$  & 0.9390 & $63.23_{-6.49}^{+7.51}$ \\
       &       & 0.0663 & $62.15_{-0.04}^{+0.03}$  & 0.1372 & $60.58_{-0.02}^{+0.02}$  & 0.0919 & $60.44_{-0.04}^{+0.03}$  & 0.0222 & $79.71_{-0.89}^{+1.01}$ \\
       &       & 0.0471 & $60.46_{-0.02}^{+0.02}$  & 0.0949 & $62.18_{-0.02}^{+0.02}$  & 0.0859 & $60.61_{-0.04}^{+0.14}$  & 0.0085 & $50.46_{-0.29}^{+0.32}$ \\
\cline{3-10}
       &   81  & 0.7484 & $62.08_{-0.02}^{+0.02}$  & 0.1624 & $62.06_{-0.02}^{+0.02}$  & 0.0579 & $60.58_{-0.02}^{+0.03}$  & 0.8786 & $74.87_{-10.2}^{+10.7}$ \\
       &       & 0.0917 & $62.19_{-0.02}^{+0.04}$  & 0.1291 & $67.37_{-0.02}^{+0.02}$  & 0.0543 & $62.06_{-0.04}^{+0.02}$  & 0.0387 & $94.17_{-0.76}^{+0.84}$ \\
       &       & 0.0847 & $60.58_{-0.02}^{+0.02}$  & 0.0698 & $58.27_{-0.02}^{+0.02}$  & 0.0516 & $58.27_{-0.02}^{+0.02}$  & 0.0136 & $96.01_{-0.29}^{+0.20}$ \\
\cline{3-10}
       &   28  & 0.9119 & $62.06_{-0.02}^{+0.02}$  & 0.4170 & $62.07_{-0.04}^{+0.03}$  & 0.1485 & $62.06_{-0.04}^{+0.03}$  & 0.8200 & $61.16_{-8.33}^{+11.8}$ \\
       &       & 0.0426 & $59.46_{-0.02}^{+0.02}$  & 0.0865 & $55.23_{-0.02}^{+0.02}$  & 0.0465 & $55.22_{-0.03}^{+0.08}$  & 0.0923 & $84.60_{-2.54}^{+3.02}$ \\
       &       & 0.0103 & $57.46_{-0.02}^{+0.02}$  & 0.0601 & $59.47_{-0.03}^{+0.02}$  & 0.0318 & $66.36_{-0.02}^{+0.02}$  & 0.0063 & $94.59_{-0.23}^{+0.25}$ \\
\hline
   6   &  333  & 0.8775 & $62.07_{-0.03}^{+0.02}$  & 0.1510 & $62.06_{-0.02}^{+0.02}$  & 0.3745 & $58.78_{-0.91}^{+0.78}$  & 0.8689 & $62.62_{-3.53}^{+4.22}$ \\
(22.78)&       & 0.0883 & $62.20_{-0.07}^{+0.14}$  & 0.0554 & $63.50_{-0.02}^{+0.02}$  & 0.0409 & $60.26_{-0.15}^{+0.04}$  & 0.0621 & $72.12_{-1.24}^{+2.02}$ \\
       &       & 0.0228 & $62.53_{-0.02}^{+0.02}$  & 0.0515 & $60.61_{-0.02}^{+0.02}$  & 0.0238 & $60.44_{-0.02}^{+0.03}$  & 0.0492 & $55.82_{-0.85}^{+0.57}$ \\
\cline{3-10}
       &   35  & 0.7534 & $62.07_{-0.02}^{+0.02}$  & 0.2041 & $62.08_{-0.04}^{+0.10}$  & 0.0434 & $70.91_{-0.02}^{+0.02}$  & 0.8626 & $59.10_{-6.85}^{+10.7}$ \\
       &       & 0.0803 & $62.17_{-0.02}^{+0.02}$  & 0.1047 & $60.72_{-0.12}^{+0.03}$  & 0.0406 & $76.59_{-0.04}^{+0.03}$  & 0.0386 & $80.33_{-1.02}^{+1.34}$ \\
       &       & 0.0798 & $60.72_{-0.02}^{+0.02}$  & 0.0649 & $65.70_{-0.02}^{+0.03}$  & 0.0304 & $72.50_{-0.02}^{+0.02}$  & 0.0209 & $86.18_{-0.93}^{+0.91}$ \\
\cline{3-10}
       &   22  & 0.7403 & $62.07_{-0.02}^{+0.02}$  & 0.1675 & $62.07_{-0.03}^{+0.02}$  & 0.0715 & $62.08_{-0.06}^{+0.10}$  & 0.7925 & $59.14_{-7.06}^{+10.9}$ \\
       &       & 0.0954 & $67.47_{-0.02}^{+0.02}$  & 0.0836 & $59.46_{-0.11}^{+0.03}$  & 0.0566 & $57.52_{-0.10}^{+0.06}$  & 0.1309 & $82.59_{-3.75}^{+4.10}$ \\
       &       & 0.0413 & $57.42_{-0.02}^{+0.02}$  & 0.0674 & $55.22_{-0.02}^{+0.02}$  & 0.0443 & $55.22_{-0.08}^{+0.02}$  & 0.0143 & $93.05_{-0.51}^{+0.59}$ \\
\cline{3-10}
       &   11  & 0.9340 & $57.47_{-0.02}^{+0.02}$  & 0.2283 & $57.45_{-0.02}^{+0.02}$  & 0.0470 & $57.44_{-0.03}^{+0.04}$  & 0.7396 & $57.55_{-5.91}^{+9.60}$ \\
       &       & 0.0649 & $64.87_{-0.02}^{+0.02}$  & 0.1457 & $64.88_{-0.04}^{+0.03}$  & 0.0421 & $64.88_{-0.02}^{+0.02}$  & 0.1427 & $78.61_{-3.51}^{+4.08}$ \\
       &       & 0.0009 & $59.87_{-0.02}^{+0.02}$  & 0.0529 & $62.05_{-0.02}^{+0.02}$  & 0.0380 & $63.13_{-0.02}^{+0.02}$  & 0.0356 & $86.64_{-1.32}^{+1.27}$ \\
\hline
   4   &  257  & 0.8109 & $62.07_{-0.02}^{+0.02}$  & 0.1354 & $62.05_{-0.02}^{+0.02}$  & 0.4164 & $56.40_{-1.23}^{+1.06}$  & 0.9079 & $67.07_{-6.02}^{+6.38}$ \\
(18.20)&       & 0.0346 & $60.60_{-0.04}^{+0.03}$  & 0.0658 & $59.40_{-0.02}^{+0.02}$  & 0.2561 & $58.78_{-0.64}^{+0.60}$  & 0.0504 & $80.75_{-1.12}^{+1.70}$ \\
       &       & 0.0244 & $60.87_{-0.02}^{+0.02}$  & 0.0428 & $60.60_{-0.02}^{+0.02}$  & 0.0682 & $54.00_{-0.35}^{+0.37}$  & 0.0122 & $54.09_{-0.52}^{+0.46}$ \\
\cline{3-10}
       &   51  & 0.4769 & $62.08_{-0.02}^{+0.02}$  & 0.2376 & $62.05_{-0.02}^{+0.02}$  & 0.1018 & $62.11_{-0.09}^{+0.15}$  & 0.9063 & $60.93_{-7.55}^{+9.78}$ \\
       &       & 0.3750 & $59.34_{-0.02}^{+0.03}$  & 0.1386 & $59.46_{-0.02}^{+0.03}$  & 0.0907 & $60.71_{-0.14}^{+0.13}$  & 0.0517 & $81.76_{-1.75}^{+2.29}$ \\
       &       & 0.1101 & $60.86_{-0.04}^{+0.03}$  & 0.1089 & $76.50_{-0.02}^{+0.02}$  & 0.0730 & $59.21_{-0.15}^{+0.13}$  & 0.0027 & $85.54_{-0.19}^{+0.14}$ \\
\cline{3-10}
       &   30  & 0.9354 & $62.05_{-0.03}^{+0.05}$  & 0.1869 & $62.06_{-0.04}^{+0.02}$  & 0.0457 & $57.44_{-0.44}^{+0.12}$  & 0.8192 & $60.03_{-7.57}^{+11.4}$ \\
       &       & 0.0212 & $60.80_{-0.08}^{+0.08}$  & 0.0737 & $62.17_{-0.02}^{+0.02}$  & 0.0305 & $76.58_{-0.04}^{+0.03}$  & 0.1112 & $83.84_{-3.06}^{+3.80}$ \\
       &       & 0.0053 & $60.59_{-0.02}^{+0.02}$  & 0.0634 & $63.57_{-0.02}^{+0.02}$  & 0.0172 & $74.80_{-0.02}^{+0.02}$  & 0.0156 & $90.54_{-0.82}^{+0.59}$ \\
\cline{3-10}
       &   19  & 0.7021 & $67.46_{-0.02}^{+0.02}$  & 0.2500 & $62.07_{-0.02}^{+0.02}$  & 0.0896 & $62.07_{-0.03}^{+0.10}$  & 0.6682 & $54.27_{-3.36}^{+6.09}$ \\
       &       & 0.2775 & $62.06_{-0.02}^{+0.02}$  & 0.1272 & $67.46_{-0.02}^{+0.02}$  & 0.0549 & $55.21_{-0.08}^{+0.07}$  & 0.2381 & $71.14_{-4.77}^{+6.32}$ \\
       &       & 0.0060 & $67.51_{-0.02}^{+0.02}$  & 0.0795 & $55.22_{-0.02}^{+0.02}$  & 0.0355 & $67.46_{-0.02}^{+0.02}$  & 0.0185 & $84.83_{-0.71}^{+0.86}$ \\
\hline
\end{tabular}
\end{center}
}
\end{table*}

\begin{table*}
\caption{%
Results of the Monte Carlo exercises with three red-giant stars
observed by the \kep\ satellite.  The quantities shown have the same
meaning as in Table~\ref{tab:error_table_model1}.
\label{tab:error_table_obs}
}
\renewcommand{\arraystretch}{1.5}
{\footnotesize
\begin{center}
\begin{tabular}{ccccccccccc}
\hline
            & \multicolumn{10}{c}{$\delta\nu$\,($\sigma = 0.022\mu$Hz)} \\
\cline{2-11}
  KIC id    & \multicolumn{2}{c}{$1\sigma$}     & \multicolumn{2}{c}{$2\sigma$}     & \multicolumn{2}{c}{$3\sigma$}     & \multicolumn{2}{c}{$4\sigma$}     & \multicolumn{2}{c}{$5\sigma$}        \\
\cline{2-11}
            & $p$ & \dpg\,(s)  & $p$ & \dpg\,(s)  & $p$ & \dpg\,(s)  & $p$ & \dpg\,(s)  & $p$ & \dpg\,(s)   \\   
\hline
            &  0.8087 & $81.54_{-0.04}^{+0.06}$ &  0.6339 & $81.53_{-0.05}^{+0.06}$ &  0.5506 & $81.52_{-0.05}^{+0.06}$ &  0.4370 & $81.52_{-0.06}^{+0.07}$ &  0.3122 & $81.53_{-0.07}^{+0.08}$ \\ 
  10200377  &  0.1169 & $82.51_{-0.02}^{+0.02}$ &  0.1993 & $82.52_{-0.04}^{+0.03}$ &  0.1757 & $82.51_{-0.08}^{+0.06}$ &  0.1976 & $84.84_{-0.04}^{+0.05}$ &  0.2658 & $84.86_{-0.05}^{+0.06}$ \\ 
            &  0.0190 & $81.69_{-0.03}^{+0.04}$ &  0.0558 & $83.42_{-0.03}^{+0.02}$ &  0.1006 & $84.83_{-0.03}^{+0.04}$ &  0.1303 & $82.49_{-0.11}^{+0.08}$ &  0.0889 & $82.47_{-0.10}^{+0.09}$ \\ 
\hline
            &  0.5389 & $76.98_{-0.02}^{+0.04}$ &  0.3463 & $76.99_{-0.04}^{+0.09}$ &  0.2796 & $79.81_{-0.17}^{+0.07}$ &  0.2495 & $79.77_{-0.15}^{+0.10}$ &  0.2206 & $81.91_{-0.11}^{+0.08}$ \\ 
   9145955  &  0.3022 & $77.72_{-0.04}^{+0.03}$ &  0.3262 & $77.72_{-0.05}^{+0.06}$ &  0.1745 & $77.72_{-0.06}^{+0.07}$ &  0.2033 & $81.92_{-0.09}^{+0.07}$ &  0.1929 & $79.74_{-0.13}^{+0.13}$ \\ 
            &  0.0708 & $76.42_{-0.03}^{+0.02}$ &  0.1434 & $79.85_{-0.06}^{+0.03}$ &  0.1645 & $77.00_{-0.06}^{+0.09}$ &  0.1759 & $78.87_{-0.12}^{+0.22}$ &  0.1307 & $78.91_{-0.19}^{+0.17}$ \\ 
\hline
            &  0.4902 & $68.49_{-0.05}^{+0.04}$ &  0.0803 & $68.48_{-0.05}^{+0.06}$ &  0.4771 & $38.37_{-5.99}^{+8.24}$ & \multicolumn{2}{c}{\blank}         & \multicolumn{2}{c}{\blank}        \\
   5866737  &  0.0807 & $72.59_{-0.03}^{+0.04}$ &  0.0371 & $72.61_{-0.05}^{+0.06}$ &  0.0165 & $68.48_{-0.05}^{+0.06}$ & \multicolumn{2}{c}{\blank}         & \multicolumn{2}{c}{\blank}        \\
            &  0.0564 & $68.76_{-0.03}^{+0.03}$ &  0.0298 & $68.78_{-0.05}^{+0.07}$ &  0.0087 & $64.79_{-0.05}^{+0.05}$ & \multicolumn{2}{c}{\blank}         & \multicolumn{2}{c}{\blank}        \\
\hline
\end{tabular}
\end{center}
}
\end{table*}

\end{document}